\documentclass[twocolumn,trackchanges,twocolappendix]{aastex631}
\usepackage{booktabs}
\usepackage{CJK}
\usepackage{multirow}
\usepackage{makecell}
\usepackage{graphicx}

\newcommand{\kms}{km~s$^{-1}$}
\newcommand{\jybeamkm}{Jy beam$^{-1}$ km$^{-1}$}
\newcommand{\jybeam}{Jy beam$^{-1}$}
\newcommand{\mjybeam}{mJy beam$^{-1}$}

\graphicspath{{./}{figures/}}
\begin{document}

\title{A low-mass line-rich core found in Massive Star-forming Region IRAS 16351-4722}

\shorttitle{low-mass line-rich core in I16351}
\shortauthors{Liu et al.}



\author[0000-0002-5789-7504]{Meizhu Liu}
\affiliation{School of Physics and Astronomy, Yunnan University, Kunming 650091, People's Republic of China}

\author[0000-0003-2302-0613]{Sheng-Li Qin}
\affiliation{School of Physics and Astronomy, Yunnan University, Kunming 650091, People's Republic of China}

\author[0000-0002-5286-2564]{Tie Liu}
\affiliation{Shanghai Astronomical Observatory, Chinese Academy of Sciences, 80 Nandan Road, Shanghai 200030, People’s Republic of China}

\author[0000-0001-9160-2944]{Mengyao Tang}
\affiliation{Institute of Astrophysics, School of Physics and Electronic Science, Chuxiong Normal University, Chuxiong 675000, People’s Republic of China}

\author[0000-0003-4603-7119]{Sheng-Yuan Liu}
\affiliation{Institute of Astronomy and Astrophysics, Academia Sinica, 11F of Astronomy-Mathematics Building, AS/NTU No. 1, Section 4, Roosevelt Road., Taipei 10617, Taiwan}

\author[0009-0009-8154-4205]{Li Chen}
\affiliation{School of Physics and Astronomy, Yunnan University, Kunming 650091, People's Republic of China}

\author[0000-0001-5710-6509]{ChuanShou Li}
\affiliation{School of Physics and Astronomy, Yunnan University, Kunming 650091, People's Republic of China}

\author[0000-0001-8277-1367]{HongQiong Shi}
\affiliation{School of Physics and Astronomy, Yunnan University, Kunming 650091, People's Republic of China}

\author[0000-0003-2090-5416]{Xiaohu Li}
\affiliation{Xinjiang Astronomical Observatory, Chinese Academy of Sciences, Urumqi, China}

\author[0000-0002-1466-3484]{Tianwei Zhang}
\affiliation{I. Physikalisches Institut, Universit{\"a}t zu K{\"o}ln, Z{\"u}lpicher Stra{\ss}e 77, 50937 K{\"o}ln, Germany}

\author[0000-0002-8149-8546]{Ken’ichi Tatematsu}
\affiliation{Nobeyama Radio Observatory, National Astronomical Observatory of Japan, National Institutes of Natural Sciences, Nobeyama, Minamimaki, Minamisaku, Nagano 384-1305, Japan}

\author[0000-0001-5950-1932]{Fengwei Xu}
\affiliation{Kavli Institute for Astronomy and Astrophysics, Peking University, 5 Yi-heyuan Road, Haidian District, Beijing 100871, China}
\affiliation{Department of Astronomy, School of Physics, Peking University, Beijing 100871, People’s Republic of China}

\author[0000-0002-5076-7520]{Yuefang Wu}
\affiliation{Department of Astronomy, School of Physics, Peking University, Beijing 100871, People’s Republic of China}

\correspondingauthor{Meizhu Liu}
\email{lmz@mail.ynu.edu.cn}

\begin{abstract}

We present ALMA sub-arcsecond-resolution observations of both continuum and molecular lines at 345 GHz towards the massive star-forming region IRAS 16351-4722 (hereafter I16351). A total of 12 dust cores were detected based on high spatial resolution observations of the continuum. Among them, a high-mass core (11.6 M$_{\odot}$) and a low-mass core (1.7 M$_{\odot}$) show abundant molecular line emissions. 164 molecular transitions from 29 species and 104 molecular transitions from 25 species are identified in the high-mass and low-mass cores, respectively. Complex organic molecules (COMs) such as CH$_3$OH, CH$_3$OCHO, CH$_3$OCH$_3$, C$_2$H$_5$OH, and C$_2$H$_5$CN are detected in the two cores. Under the assumption of local thermodynamic equilibrium (LTE), rotational temperatures and column densities of the COMs are derived with the XCLASS software. The maximum rotation temperature values in the low-mass core and the high-mass core were found to be approximately 130 K and 198 K, respectively. Additionally, the line widths in the high-mass core are larger than those in the low-mass one. Abundant complex organic molecular line transitions, high gas temperatures, and smaller line widths indicate the presence of a low-mass line-rich core in the massive star formation region for the first time, while the high-mass line-rich core shows hot core property. When comparing the molecular abundances of CH$_3$OH, CH$_3$OCHO, CH$_3$OCH$_3$ and C$_2$H$_5$OH of the two cores with other hot cores and hot corinos reported in the literature, we further confirm that both a hot core and a low-mass line-rich core are simultaneously detected in I16351.

\end{abstract}


\section{Introduction} \label{sec:intro}

The development of radio telescopes has facilitated in-depth investigations of interstellar molecules \citep[e.g.,][] {2018A&A...620L...6T,2019A&A...628A..27B,2019A&A...624L...5P}. So far, about 300 molecules \footnote{\url{https://cdms.astro.uni-koeln.de/classic/molecules}} composed of 16 different elements, with atomic numbers ranging from 2 to 70, have been detected in interstellar space. Complex organic molecules (COMs), which consist of six or more atoms containing carbon atoms, serve as valuable indicators of the physical conditions and history of their sources through their spectra and chemical properties \citep{2009ARA&A..47..427H}. COMs are mostly detected in hot cores and hot corinos \citep{2020ARA&A..58..727J}, but also on the surface of the embedded disks where the temperature is warm ($\gtrsim$ 100 K; \citep[e.g.,][]{2017ApJ...843...27L}). Jets and outflows can also produce the emission of COMs by sputtering and/or shock chemical events \citep[e.g.,][]{2008ApJ...681L..21A, 2017MNRAS.469L..73L}.

The hot core is characterized by a dense ($\gtrsim$ 10$^7$ cm$^{-3}$) region on scales $<$ 0.1 pc, abundant molecular line emissions with high gas temperatures ($\gtrsim$ 100 K) and high molecular abundances \citep{2000prpl.conf..299K,2009ARA&A..47..427H}. Hot core observations in massive star-forming regions have been promoted by the millimetre/submillimetre interferometric arrays greatly \citep{2002ApJ...576..255L,2004ApJ...617..384R,2010ApJ...711..399Q,2015ApJ...803...39Q,2005ApJ...628..800B,2014A&A...569A..11S,2013ApJ...775L..31S,2018ApJ...857...35S,2014ApJ...786...38H,2016A&A...587A..91B,2019A&A...628A..10B,2016MNRAS.455.1428R,2017A&A...604A..60B,2017A&A...604A..32P,2017A&A...598A..59R,2018A&A...618A..46A,2018ApJS..236...45G,2018A&A...620L...6T,2019A&A...628A...2B,2019A&A...632A..57C,2019ApJ...871..112X,2020A&A...636A.118M,2020ApJ...901...37L,2020ApJ...898...54T,2021ApJ...909..214L,2021A&A...655A..86V}. Hot corino sources are characterized by localized zones that surround low-mass to intermediate-mass protostars, and they have warm gas temperatures ($\sim$ 100 K), and compact COM emissions ($\sim$ 100 AU) \citep{2004ASPC..323..195C}. Since the first proposal of the definition of ``hot corino" \citep{2004ASPC..323..195C}, only about a couple of tens of hot corino sources have been discovered \citep{2020ApJ...898..107H}. Recently, \citet{2020A&A...635A.198B} observed 26 solar-type star-forming regions, in which warm methanol is detected in 12 sources. \citet{2020ApJ...898..107H,2022ApJ...927..218H} reported the detection of 11 hot corino sources among 56 Class 0/I protostellar cores from the ALMA observations of Orion Planck Galactic Cold Clumps. In summary, high-resolution interferometer observations expand the sample size of hot corinos. 

IRAS 16351-4722 (hereafter I16351) is located at a distance of 3.02 kpc with a high clump mass of 1.58 $\times$ 10$^3$ M$_{\odot}$ and infrared bolometric luminosity of 7.9 $\times$ 10$^4$ L$_{\odot}$ \citep{2020MNRAS.496.2790L}. ALMA observations of CO (3-2) suggested outflow motion in this region \citep{2021MNRAS.507.4316B}. CO (4$-$3) line in I16351 shows a blue profile, characterized by an asymmetric parameter of $-$0.76 and a virial parameter of 1.93, indicating a global collapse within this region \citep{2021RAA....21...14Y}. These observations suggested that I16351 is an active massive star-forming region. More recently, \citet{2022MNRAS.511.3463Q} identified the emissions of C$_2$H$_5$CN, CH$_3$OCHO, and CH$_3$OH in the 3\,mm band, with rotational temperatures exceeding 100 K, implying the presence of a hot core in I16351. Therefore, I16351 is a valuable target for studying the physical and chemical conditions involved in the massive star-forming processes.

In this work, we present high spatial resolution ALMA observations toward I16351 at 345 GHz. Section\,\ref{sec:obs} introduces ALMA Band 7 observations and data reduction. We describe the observational results in Section\,\ref{sec:res}. In Section\,\ref{sec:dis}, we discuss the results of the study, and our conclusions are summarized Section\,\ref{sec:con}.

\section{Observations}\label{sec:obs}

This study utilized the ALMA Band 7 observation data (Project ID: 2017.1.00545S, Project PI: Liu Tie). The data collection took place on May 18th to 20th, 2018, during ALMA Cycle 5. A total of 43 12-meter array were deployed, using the C43-1 array configuration. The Band 7 observational data covered four spectral windows (SPWs 31, 29, 27, and 25) that encompassed the following frequency ranges: (i) 342.36$-$344.24 GHz, (ii) 344.25$-$346.09 GHz, (iii) 356.60$-$357.07 GHz, and (iv) 354.27$-$354.74 GHz. SPWs 25 and 27 have a narrow bandwidth of 468.75\,MHz and a spectral resolution of 0.24\,\kms.  SPWs 29 and 31 have a broad bandwidth of 1875\,MHz and a spectral resolution of 0.98\,\kms, which intend to cover a broad range of spectral lines. The purpose of this article is to analyze chemistry using molecular spectra, so the main analysis focuses at the broad spectral window SPWs 31 and SPWs 29. The observations were conducted using the Mosaic mode, providing an approximate field of view of 46 $\arcsec$ at 345 GHz. The phase center coordinates were R.A.(J2000) = $16^{h}38^{m}49^{s}$ and Decl.(J2000) = $-47$°$28\arcmin 03\arcsec$. Data processing was performed using the Common Astronomy Software Applications Package (CASA version 5.3.0) \citep{McMullin2007CASA}. The flux and bandpass calibration was initially applied using J1924-2914, J1427-4206, and J1517-2422. The phase calibration was carried out using J1650-5044. Imaging was carried out with \texttt{tclean} task and \texttt{robust}=0.5 was used to balance the weight between the long and short baseline data. To improve the final image, a self-calibration operation was performed on both the continuum and is applied to the solution of self-calibration from the continuum on the line data. The 870$\,\rm \mu$m continuum data were extracted from line-free channels. The final synthesized beam size is $0.8\arcsec$ $\times$ $0.7\arcsec$. The continuum RMS noise level is 1.3 \mjybeam, and the average line sensitivity is 5.05 \mjybeam per channel.

\section{Results}\label{sec:res}
\subsection{Continuum}
\begin{figure*}
\centering
\includegraphics[scale=0.75]{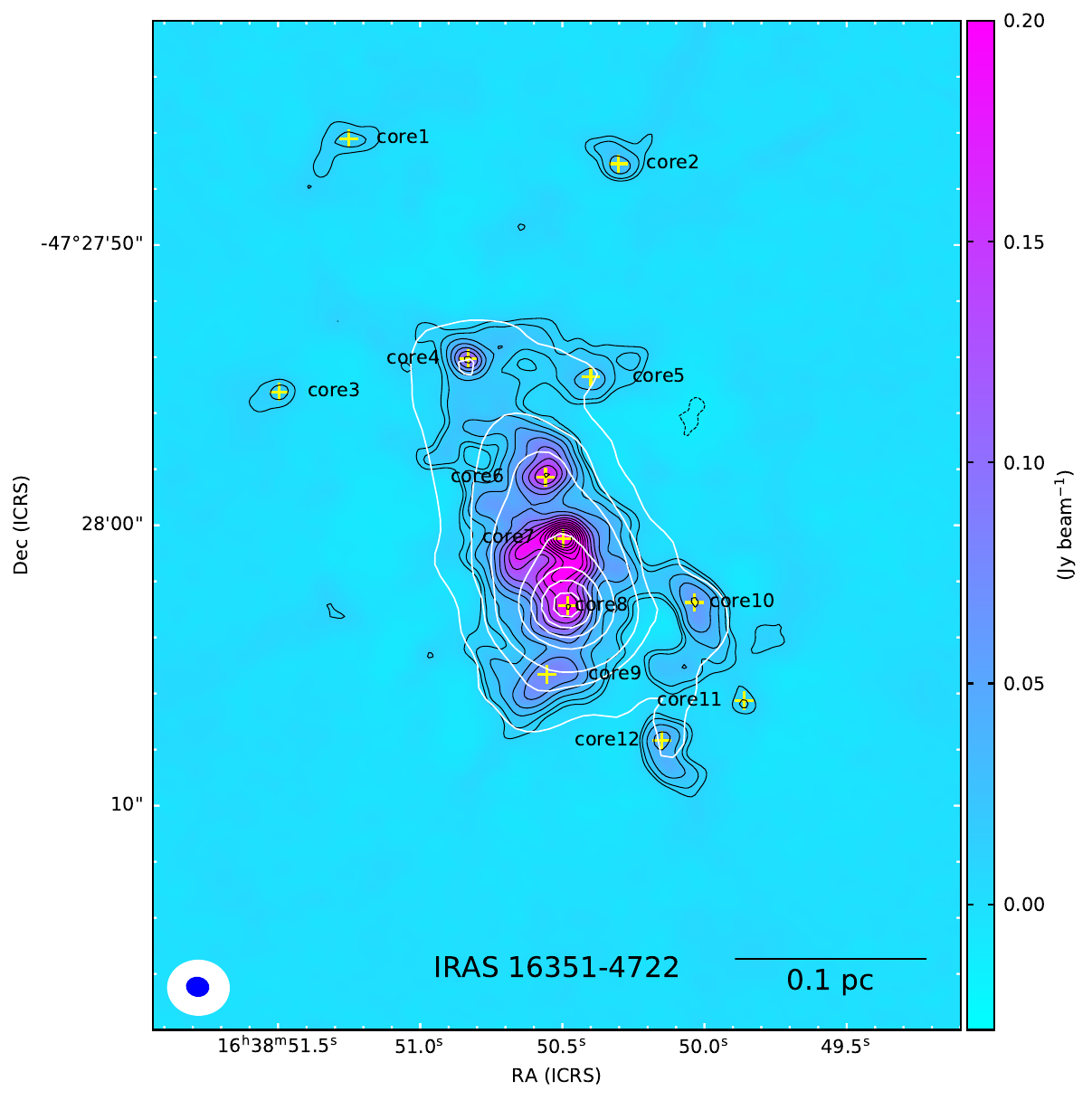}
\caption{870$\rm \,\rm \mu$m continuum emission map of I16351 from the ALMA Band 7 observations in color scale and black contours overlaid with 3 mm continuum in white contours. The synthesized beam size is shown in the bottom-left corner. The black contours start at a level of 8$\sigma$=0.0104 \jybeam and increase in the following power-law function D=4$\times$N$^p$+8, where D is the dynamical range of the intensity map as follows: 0.0104, 0.0156, 0.0267, 0.0423, 0.0614, 0.0841, 0.1099, 0.1386, 0.1701, 0.2043, 0.2411, 0.2803, 0.3219, 0.3658, 0.4119 Jy beam$^{-1}$, the p of 1.6 is the power index, N=15 is the number of contours used. The white contour levels are [3\%, 6\%, 10\%, 30\%, 50\%, 70\%, 90\%]$\times$peak integrated intensities (F$\rm _{peak}$ = 0.1193 \jybeamkm). The crosses show the peak positions of each core.}\label{fig:contin}
\end{figure*}

Previous ALMA 3\,mm observations with angular resolution
$1.2\arcsec$-$1.9\arcsec$ \citep{2020MNRAS.496.2790L} only detected one continuum core \citep{2022MNRAS.511.3463Q}. In Figure\,\ref{fig:contin}, 12 continuum cores have been detected in our ALMA high-angular continuum map. From top to bottom, we name these cores from core 1 to core 12. The continuum emission shows a complicated morphology, accompanied by a north-south extended structure. Core 7 and 8 are located in the center of the region. The brightest core is core 7 and the second is core 6. The other cores are located away from the three cores. The 870$\,\rm \mu$m continuum emission map in color scale and black contour, overlaid with 3\,mm continuum in white contour. The white contour of the 3\,mm image shows a simple morphology with a single core detected \citep{2022MNRAS.511.3463Q}. The emission peak of core 8 in this work is consistent with the peak position of the 3\,mm continuum.

We performed a two-dimensional Gaussian fitting to the 12 cores. The positions, deconvolved source sizes, peak intensities, and flux densities of these cores are listed in Table\,\ref{tab:fitcores}.
According to Table\,\ref{tab:fitcores}, the flux densities range from 33 to 3010 mJy. Additionally, the peak intensities range from 17 to 320 \mjybeam. Among all cores, core 7 has the largest flux density and peak intensity.

\begin{table*}[ht!]
\caption{Parameters of Continuum Sources}
\label{tab:fitcores}
\scalebox{0.75}{
\begin{tabular}{cccccccccc} 
\hline
\hline
Name &     R.A.    &     Decl.   & Peak Intensity   & Integrated Flux$^a$ &   Source Size$^a$     &  $\rm M_{core}$ & $\rm N_{H_2}$ & T $^b$ & Catalog \\
&    (J2000)  &    (J2000)  &    (\mjybeam)     &   (mJy)      &  ({\arcsec $\times$\arcsec})   &  $\rm M_{\odot}$  & $\rm cm^{-2}$ & K    \\
\hline
core 1 & 16:38:51.2 & -47:27:46.2 & 18(2)& 111(12)&2.4 $\times$ 1.1 & 3.0(0.7)  &7.6 $\times$ 10$^{22}$(0.8) &30(6) & Dust\\
core 2 & 16:38:50.3 & -47:27:47.1 & 34(3) & 95(10) & 1.1 $\times$ 0.8 & 2.6(0.6) & 2.0 $\times$ 10$^{23}$(0.2) & 30(6) & Dust\\
core 3 & 16:38:51.5 & -47:27:55.2 & 18(4) & 65(4) & 1.4 $\times$ 0.9 & 1.8(0.4) & 9.4 $\times$ 10$^{22}$(0.6) & 30(6) & Dust\\
core 4 & 16:38:50.8 & -47:27:54.1 & 113(3) & 220(9)& 0.8 $\times$ 0.6   & 1.7(0.2) &  2.3 $\times$ 10$^{23}$(0.09) & 90(9) & CH$_3$OCHO\\
core 5 & 16:38:50.4 & -47:27:54.7 & 35(1) & 146(4) & 1.5 $\times$ 1.0  & 4.0(0.8) & 1.8 $\times$ 10$^{23}$(0.05) & 30(6) & Dust\\
core 6 & 16:38:50.6 & -47:27:58.1 & 144(10) & 1024(79)     &    1.9 $\times$ 1.6  & 8.0(1.3) &  1.7 $\times$ 10$^{23}$(0.1)  & 87(12) & H$_2$CS\\
core 7 & 16:38:50.5 & -47:28:00.8 &  320(45)  &  3010(470)  &    2.2 $\times$ 1.9     & 11.6(1.9)  &  1.9 $\times$ 10$^{23}$(0.4) &170(10) &CH$_3$OCHO\\
core 8 & 16:38:50.5 & -47:28:02.8 & 199(4) &  1511(36)     &    2.1 $\times$ 1.6    & 11.4(1.3)   &  2.3 $\times$ 10$^{23}$(0.05)   & 91(10) &CH$_3$OCHO\\
core 9 & 16:38:50.6 & -47:28:05.3 & 79(1) &  597(10)     &    2.7 $\times$ 1.2   & 7.6(0.6)   &  1.6 $\times$ 10$^{23}$(0.03)  & 56(4) & H$_2$CS\\
core 10 & 16:38:50.0 & -47:28:02.8 & 63(4) & 378(23)& 2.0 $\times$ 0.9 & 4.6(0.6) & 1.7 $\times$ 10$^{23}$(0.1) & 58(7) & H$_2$CS\\
core 11 & 16:38:49.9 & -47:28:06.2 & 17(1)&33(1) &0.8 $\times$ 0.6 & 0.3(0.02) & 4.0 $\times$ 10$^{22}$(0.1) &78(5) &H$_2$CS\\
core 12 & 16:38:50.1 & -47:28:07.7 & 49(4) & 170(9)& 1.3 $\times$ 0.9  & 2.4(0.6) & 1.4 $\times$ 10$^{23}$(0.07)&51(12) &H$_2$CS\\
\hline
\end{tabular}
}
\tablecomments{$^a$ The total flux density and deconvolved size of each core are obtained from 2D Gaussian fitting to 870$\,\rm \mu$m continuum. \\
$^b$ The dust temperature used is the excitation temperature of  H$_2$CS or CH$_3$OCHO molecules. If there are no H$_2$CS or CH$_3$OCHO molecules in the core, we assume that the core's dust temperature is 30K \citep{2020MNRAS.496.2790L}.}
\end{table*}

To estimate the masses of the cores, we followed the procedure of \cite{1983QJRAS..24..267H} for an optically thin emission with a single temperature:

\begin{equation}
    \label{eq:core_mass}
    M_{\text{core}} = \frac{D^2 S_\nu \eta}{\kappa_\nu B_\nu (T_d)},
\end{equation}
where {S$\rm _\nu$} is the continuum integrated flux, {D} is the distance to the source, $\eta$ is the gas-to-dust ratio (100), and $\rm B _\nu (T_\mathrm{d})$ is the Plank function at the dust temperature $\rm (T_\mathrm{d})$. $\rm {\kappa_\nu}$ is the dust mass absorption coefficient of 1.89 cm$^{2}$g$^{-1}$ at 870$\,\rm \mu$m in \citet{1994A&A...291..943O}, assuming grains with thin ice mantles and a gas density of 10$^{-6}$\,cm$^{-2}$. The dust temperature is assumed to be equal to CH$_3$OCHO rotational temperatures in hot regions or H$_2$CS rotational temperatures in warm regions because H$_{2}$CS has been observed towards cold molecular clouds, protostellar cores, hot cores, circumstellar envelopes, and protoplanetary discs \citep{2022A&A...661A.111S}.

We calculate the source-averaged $\rm H_{2}$ column density according to \citet{2009A&A...504..415S}, assuming the dust emission at 870$\,\rm \mu$m is optically thin:
\begin{equation}
\textit{N(H$ _{2} $)}=\frac{{S_\nu}{\eta}}{\mu
{m_{H}{\Omega}{\kappa_{\nu}{B_{\nu}(T_d)}}}},
\end{equation}\\
where $\Omega$ is the solid angle subtended by the source, $\mu$ is the mean molecular weight of the molecular cloud, which we assume to be equal to 2.8 \citep{2008A&A...487..993K}, and {m$_{H}$} is the mass of a hydrogen atom. The mean optical depth of the continuum cores can be calculated by \citep[e.g.,][]{2010ApJ...723.1665F,2021A&A...648A..66G}:
\begin{equation}
\textit{$\tau_{870}$}=-ln[1-\frac{{S_\nu}}{\Omega B_{\nu}(T_d)}]
\end{equation}
 $\tau_{870}$ of the 12 cores is of the order of
10$^{-2}$, therefore the optically thin assumption is reasonable. From Table\,\ref{tab:fitcores}, $\rm M_{core}$ is in the range of 0.3 M$_{\odot}$ $-$ 11.6 M$_{\odot}$. $\rm N_{H_2}$ ranges from 4.0 $\times$ 10$^{22}$ cm$^{-2}$ $-$ 2.3 $\times$ 10$^{23}$ cm$^{-2}$.

\subsection{Molecular line identification}

We extract the spectral lines towards each core, and find that core 7 with a high mass of 11.6 M$_{\odot}$ and core 4 with a low mass of 1.7 M$_{\odot}$ have rich line emission when compared to the other cores. In this work, we focused on these line-rich cores for further analysis. To identify the observed molecular line transitions, we used the eXtended CASA Line Analysis Software Suite (XCLASS\footnote{\url{https://xclass.astro.uni-koeln.de}}; \citep{2017A&A...598A...7M}). XCLASS accesses the Cologne Database for Molecular Spectroscopy (CDMS\footnote{\url{http://cdms.de}}; \citep{2005JMoSt.742..215M}) and the Jet Propulsion Laboratory molecular databases (JPL\footnote{\url{http://jpl.de}}; \citep{1998JQSRT..60..883P}). We assumed that the molecular gas was in local thermodynamic equilibrium (LTE) and used the XCLASS program for the line identification. The input parameters for the modelling include the source size, line velocity width, velocity offset, rotation temperature, and column density \citep{2017A&A...598A...7M}. Figure\,\ref{fig:core} illustrates the molecular identification results of the two cores. Each plot shows the molecular line flux changing with frequency in the SPW 31 and SPW 29 spectral windows, and the molecular names are marked above the lines. From Figure\,\ref{fig:core}, one can see that the line fluxes in the high-mass core are generally much higher than those in the low-mass core. The line widths (FWHMs) of the high-mass core are broader than those of the low-mass core (see also Section \ref{sec:dis}). In total, 164 molecular line transitions from 29 species were identified in the high-mass core, while 104 molecular line transitions from 25 species were identified in the low-mass core. The common molecules detected in both the high-mass and low-mass cores include CH$_{3}$OCHO, CH$_{3}$OCHO v$_{18}$=1, CH$_{3}$OCH$_{3}$, CH$_{3}$OH, H$_{2}$CS, CH$_{3}$COCH$_{3}$, C$_{2}$H$_{5}$CN, t-HCOOH, H$_{2}$CCO, $^{13}$CH$_{3}$OH, aGg${'}$-(CH$_{2}$OH)$_{2}$, CH$_3$CHO, C$_2$H$_5$OH, HC$_3$N and SO$_2$. Formamide (NH$_2$CHO), the simplest possible amide that has been considered a potential prebiotic molecule and a potential precursor to heavier COMs, such as amino acids \citep[e.g.,][]{2015MNRAS.449.2438L,2019ESC.....3.2122L}, is also identified in both the high-mass and low-mass cores. $^{34}$SO$_2$, C$_{2}$H$_{3}$CN,
C$_{2}$H$_{3}$CN v$_{11}$=1 and CH$_{3}$C$_{4}$H are only observed in the high-mass core. CH$_3$C$_3$N, CH$_{3}$SH v$_{12}$=1 and c-H$_{2}$C$_{3}$O are only detected in the low-mass core. Therefore, the differences in molecular compositions appear to be caused by differences in their chemical properties between the high-mass and the low-mass core.

\begin{figure*}[!ht]
\newcounter{2}
\setcounter{2}{\value{figure}}
\setcounter{figure}{1}
\centering
\includegraphics[scale=0.35]{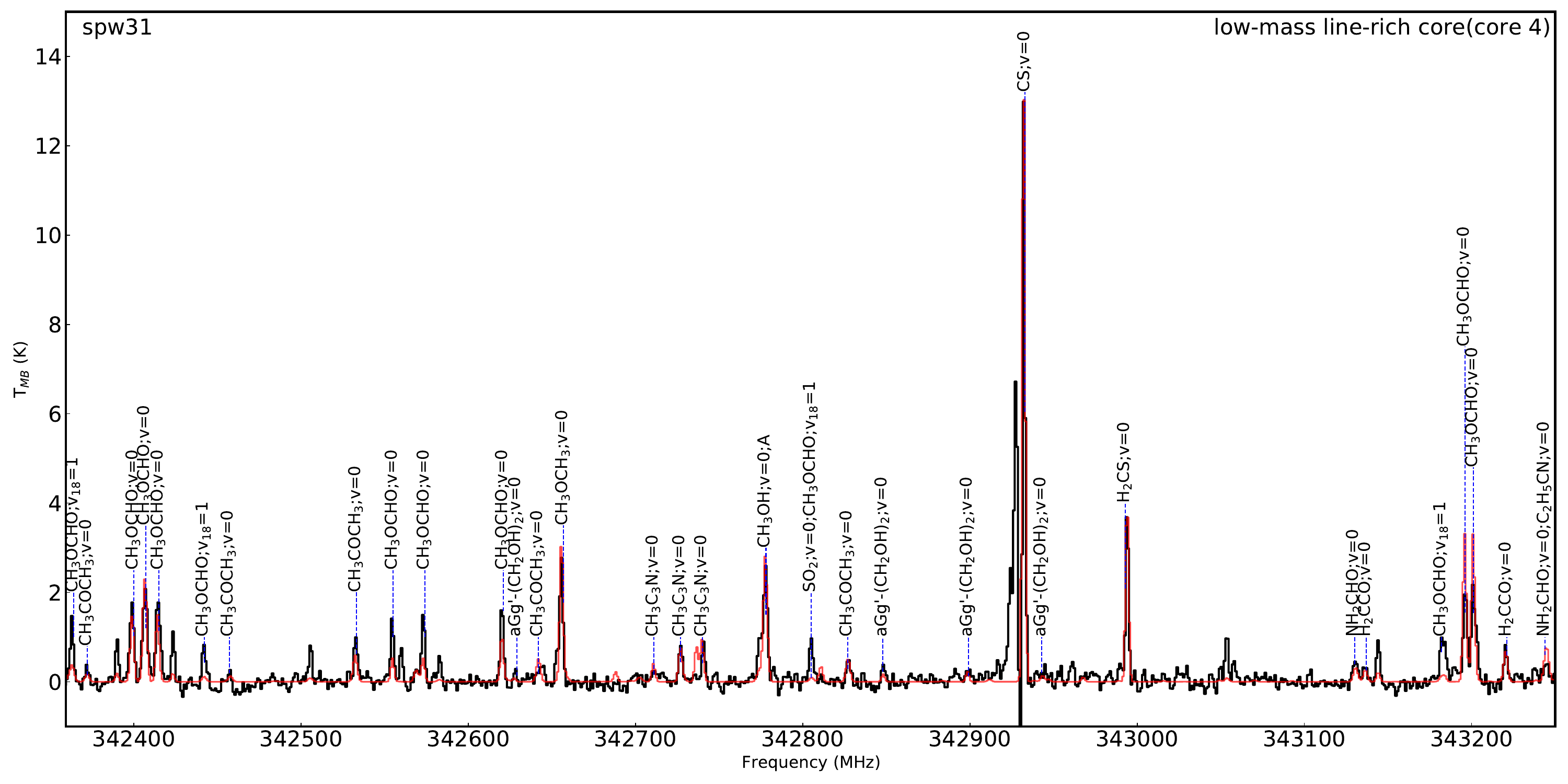}
\includegraphics[scale=0.35]{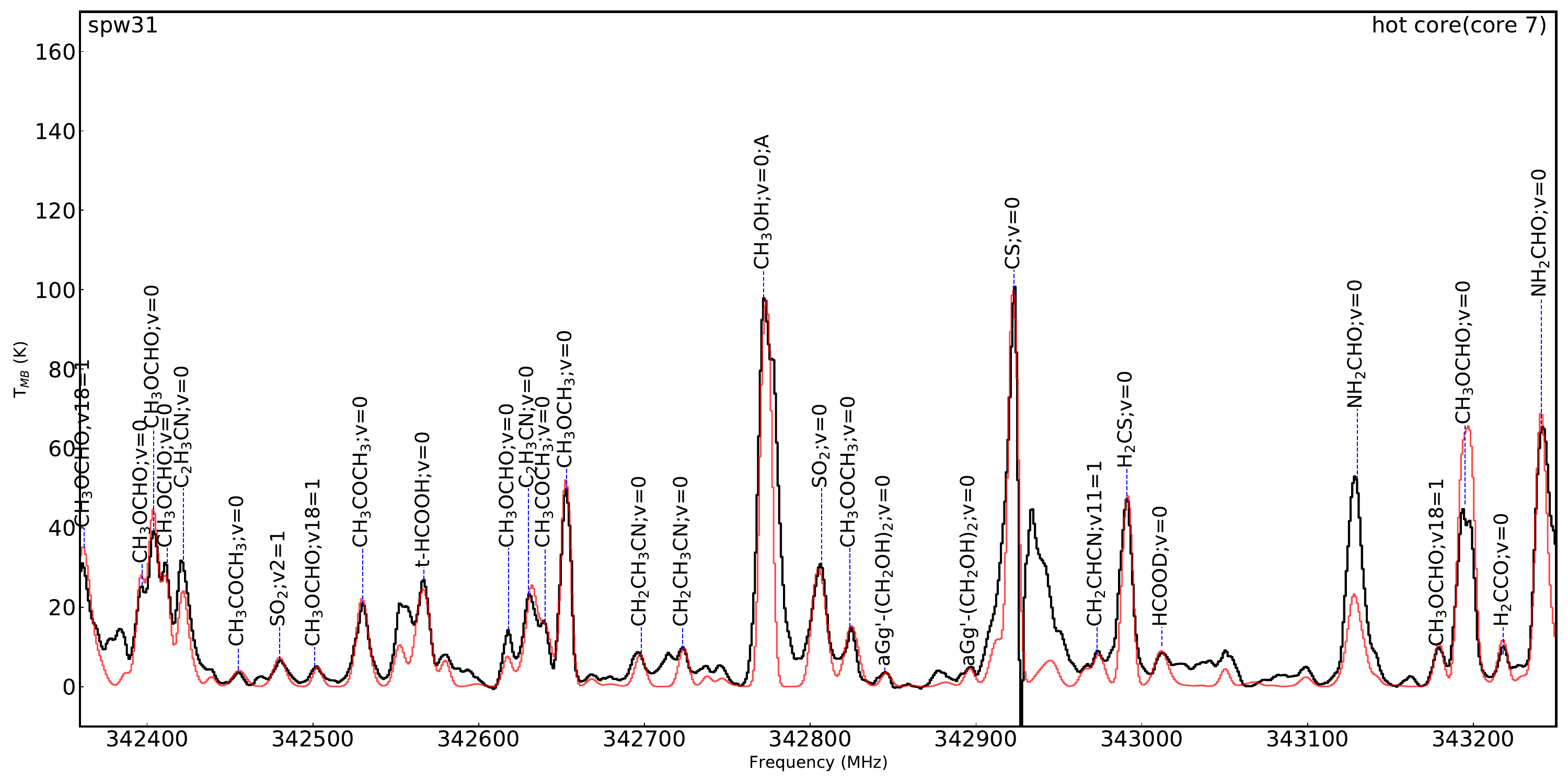}
\caption{Sample spectra in spw31 and spw29 toward the hot core and the low-mass line-rich core. The segments of spw31 and spw29 cover the frequency range from 342 to 346 GHz. The frequency scale is in terms of the rest frequency. The black curves are the observed spectra and the red curves indicate the simulated LTE spectra. }
\end{figure*}
\addtocounter{figure}{-1}

\begin{figure*}[ht!]
\setcounter{2}{\value{figure}}
\setcounter{figure}{1}
\centering
\includegraphics[scale=0.35]{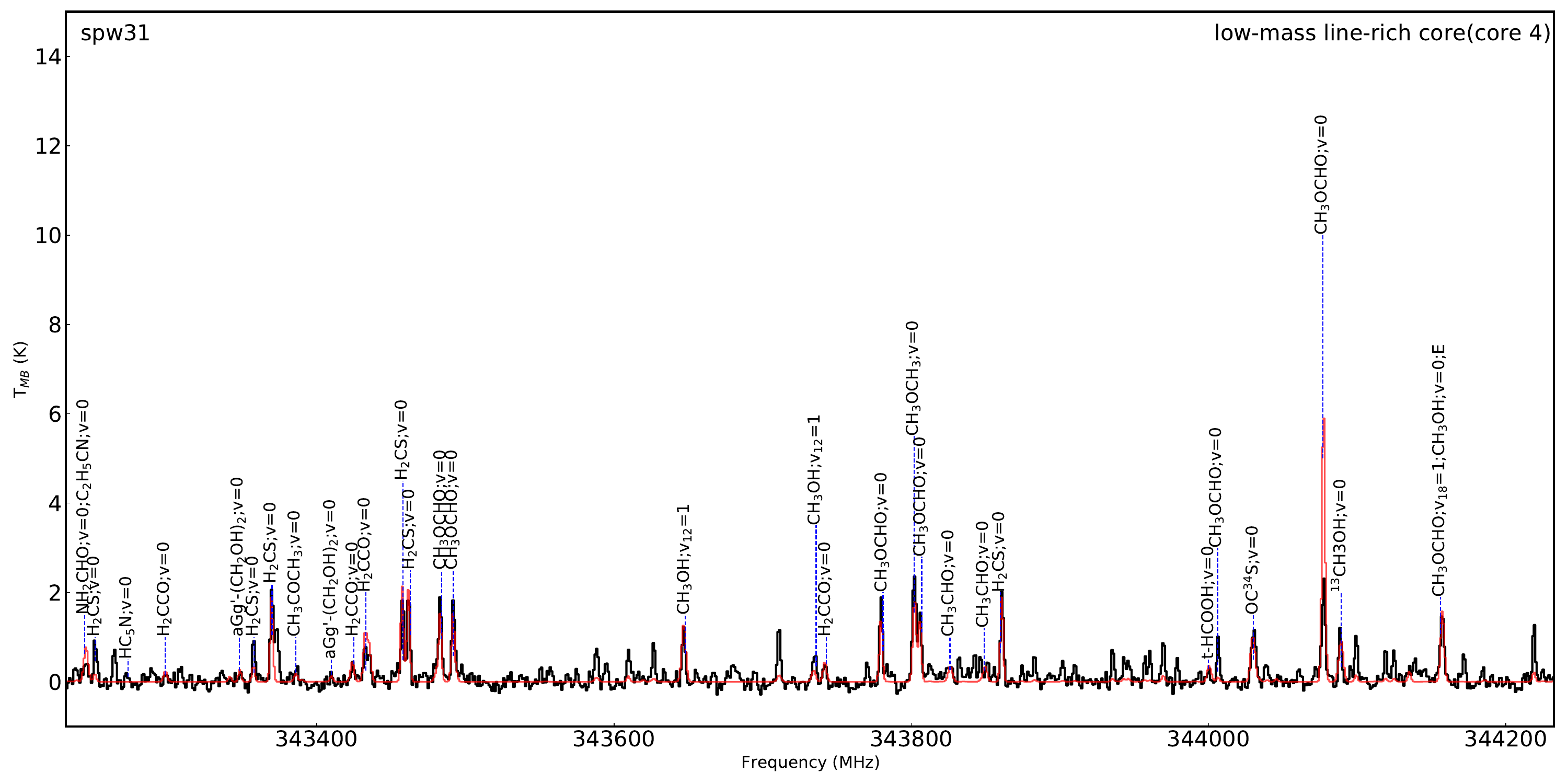}
\includegraphics[scale=0.35]{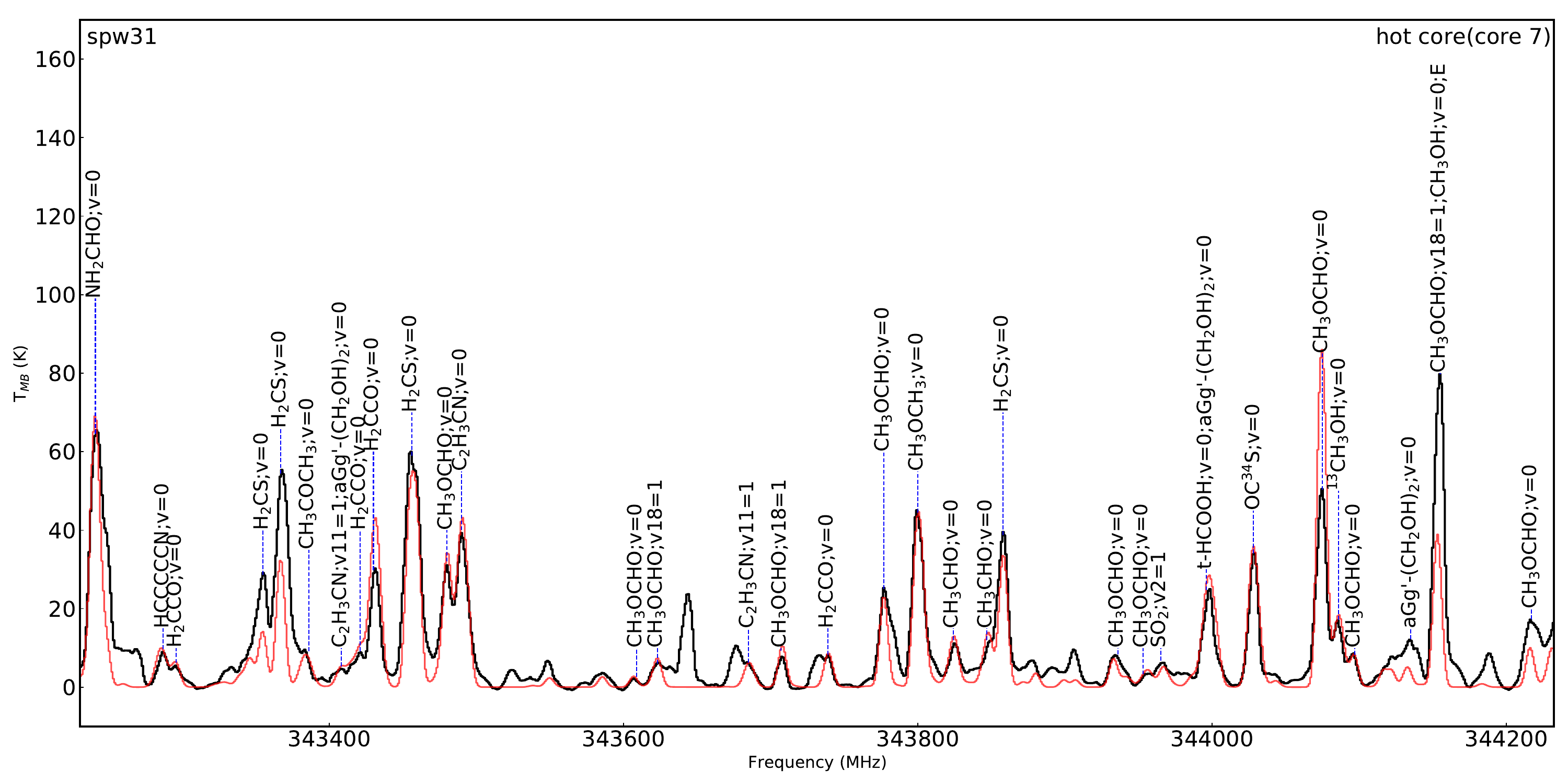}
\caption{(continued)}
\end{figure*}
\addtocounter{figure}{-1}

\begin{figure*}[ht!]
\setcounter{2}{\value{figure}}
\setcounter{figure}{1}
\centering
\includegraphics[scale=0.35]{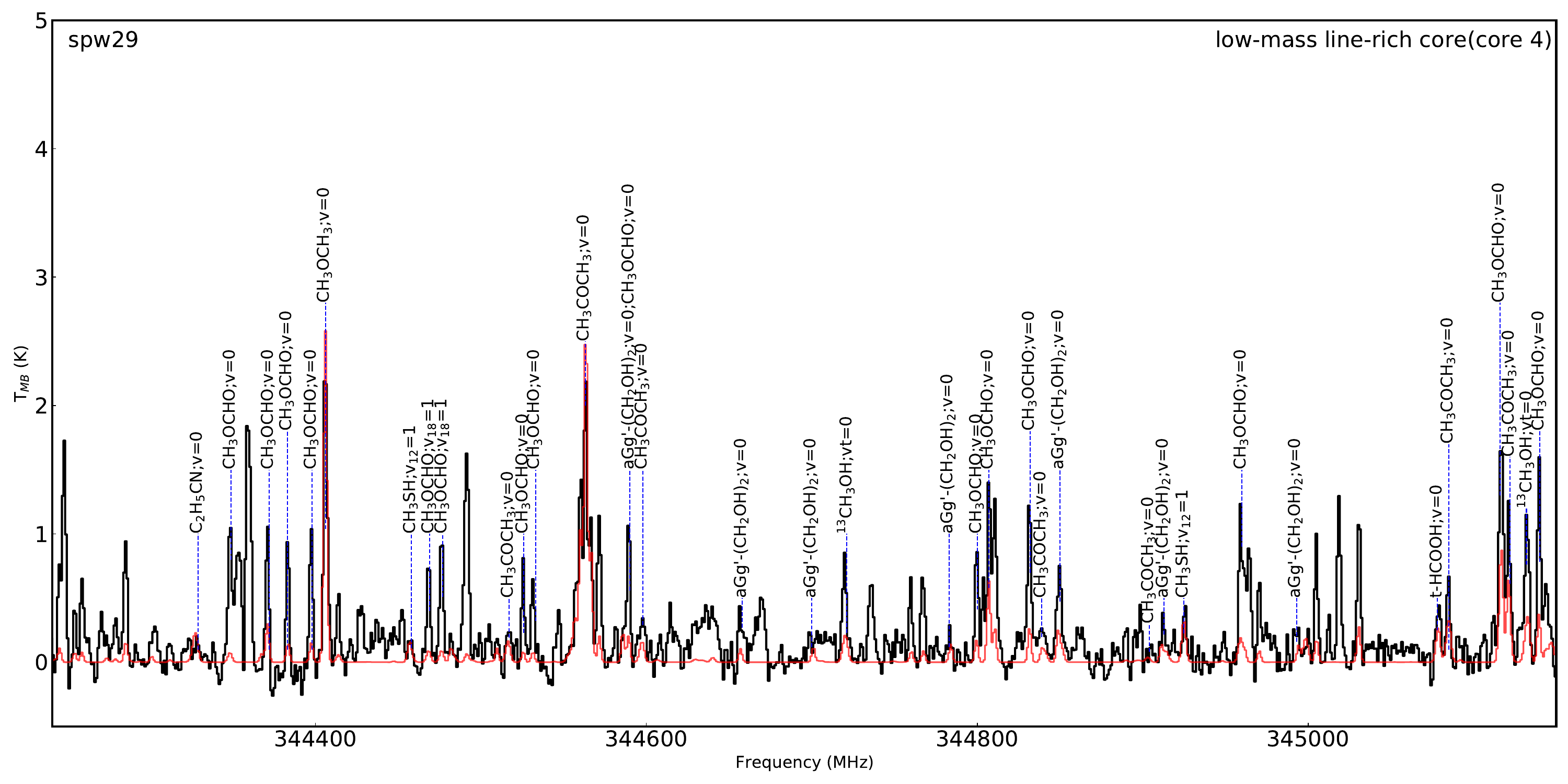}
\includegraphics[scale=0.35]{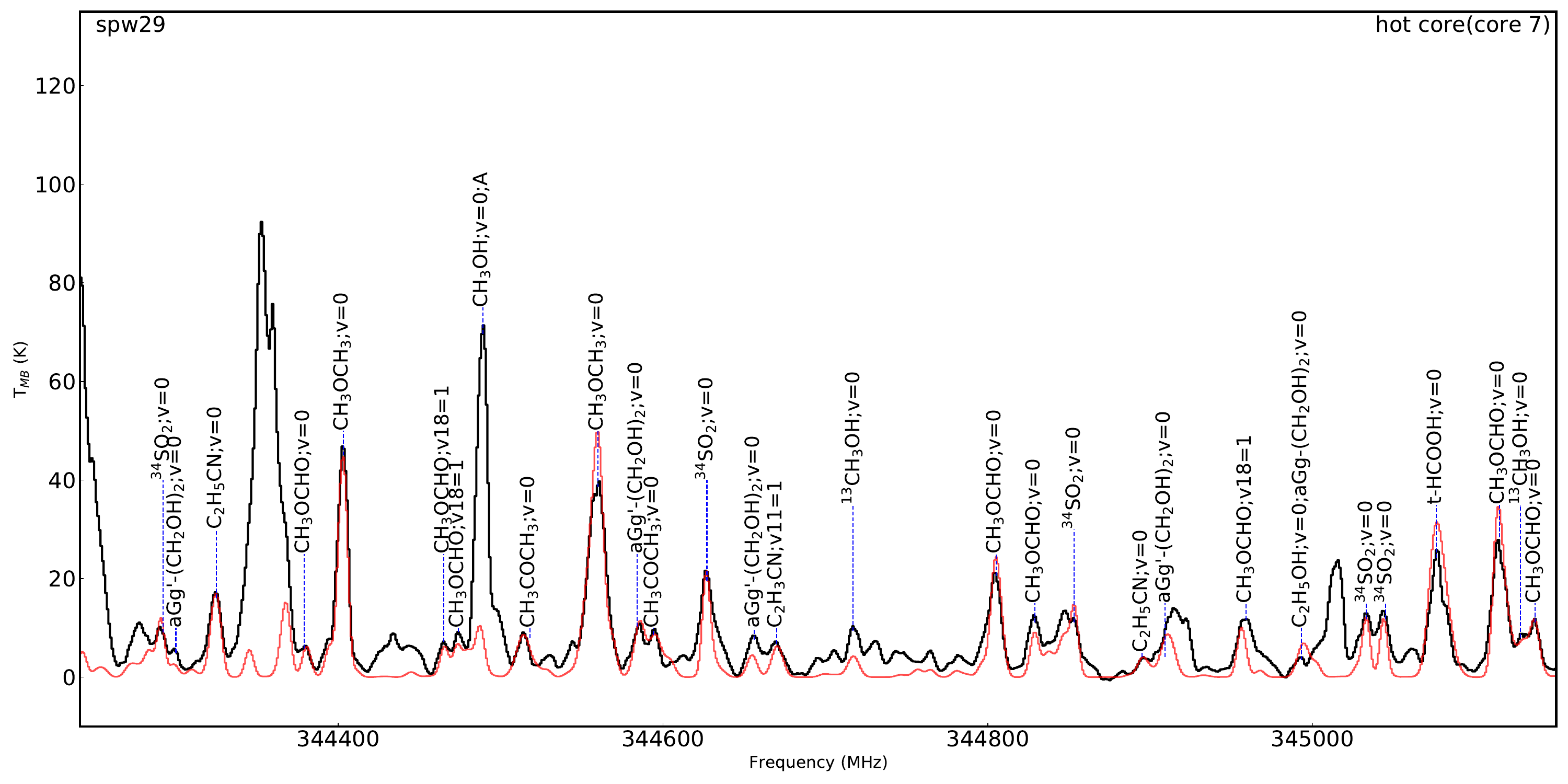}
\caption{(continued)} 
\end{figure*}
\addtocounter{figure}{-1}

\begin{figure*}[ht!]
\setcounter{2}{\value{figure}}
\setcounter{figure}{1}
\centering
\includegraphics[scale=0.35]{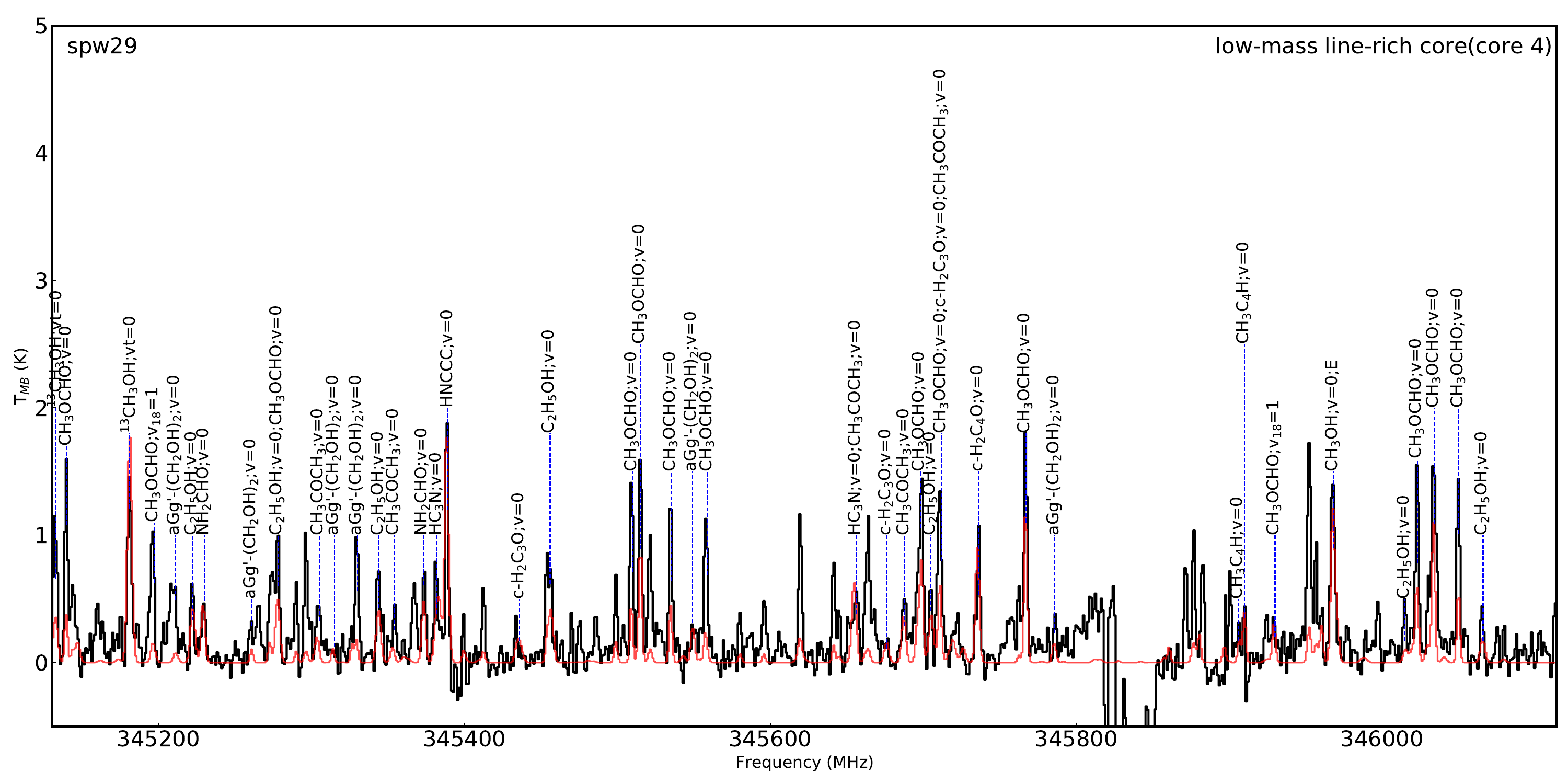}
\includegraphics[scale=0.35]{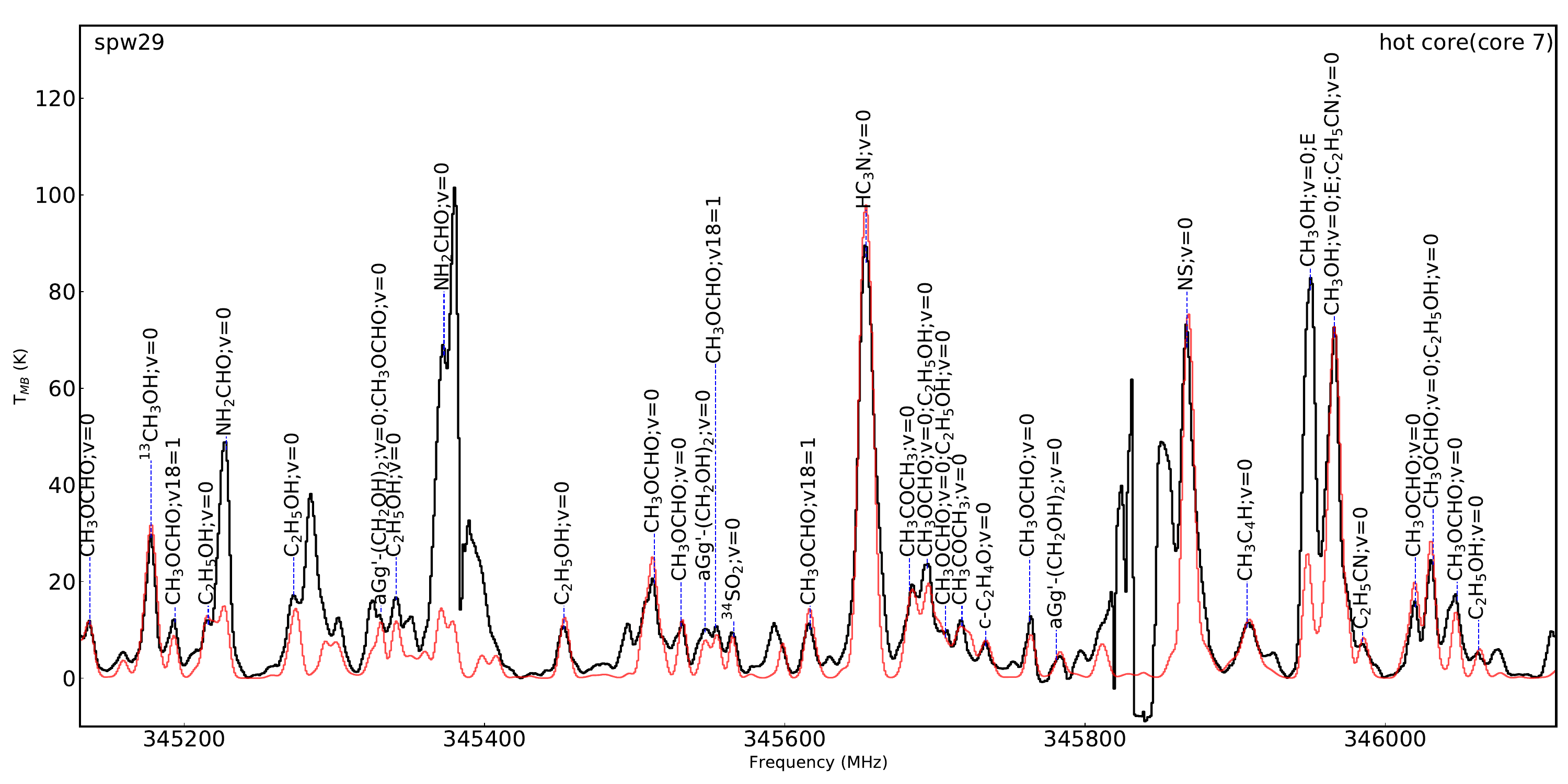}
\caption{(continued)} 
\label{fig:core}
\end{figure*}

\subsection{Gas distribution}
We have detected numerous molecules in the two cores, and selected representative molecules for imaging their spatial distributions. Figure\,\ref{fig:mom} shows the integrated intensity maps of the oxygen-, nitrogen- and sulfur-bearing molecules. Gas emissions of CH$_3$OCHO, $^{13}$CH$_3$OH, CH$_{3}$OH, CH$_{3}$COCH$_{3}$, HCOOH, CH$_{3}$CHO, C$_2$H$_5$CN, NH$_{2}$CHO, HC$_3$N, H$_{2}$CS, and SO$_{2}$ are distributed over both the low-mass and the high-mass continuum cores, while C$_2$H$_3$CN only concentrated on the high-mass core, as shown in Figure\,\ref{fig:mom}. By comparing the above 11 molecular gas distributions in both two cores, the distributions of sulfur-bearing molecules are more extended, and nitrogen-bearing molecules show compact structures.

\begin{figure*}[!ht]
\newcounter{3}
\setcounter{3}{\value{figure}}
\setcounter{figure}{2}
\centering
\includegraphics[scale=0.8]{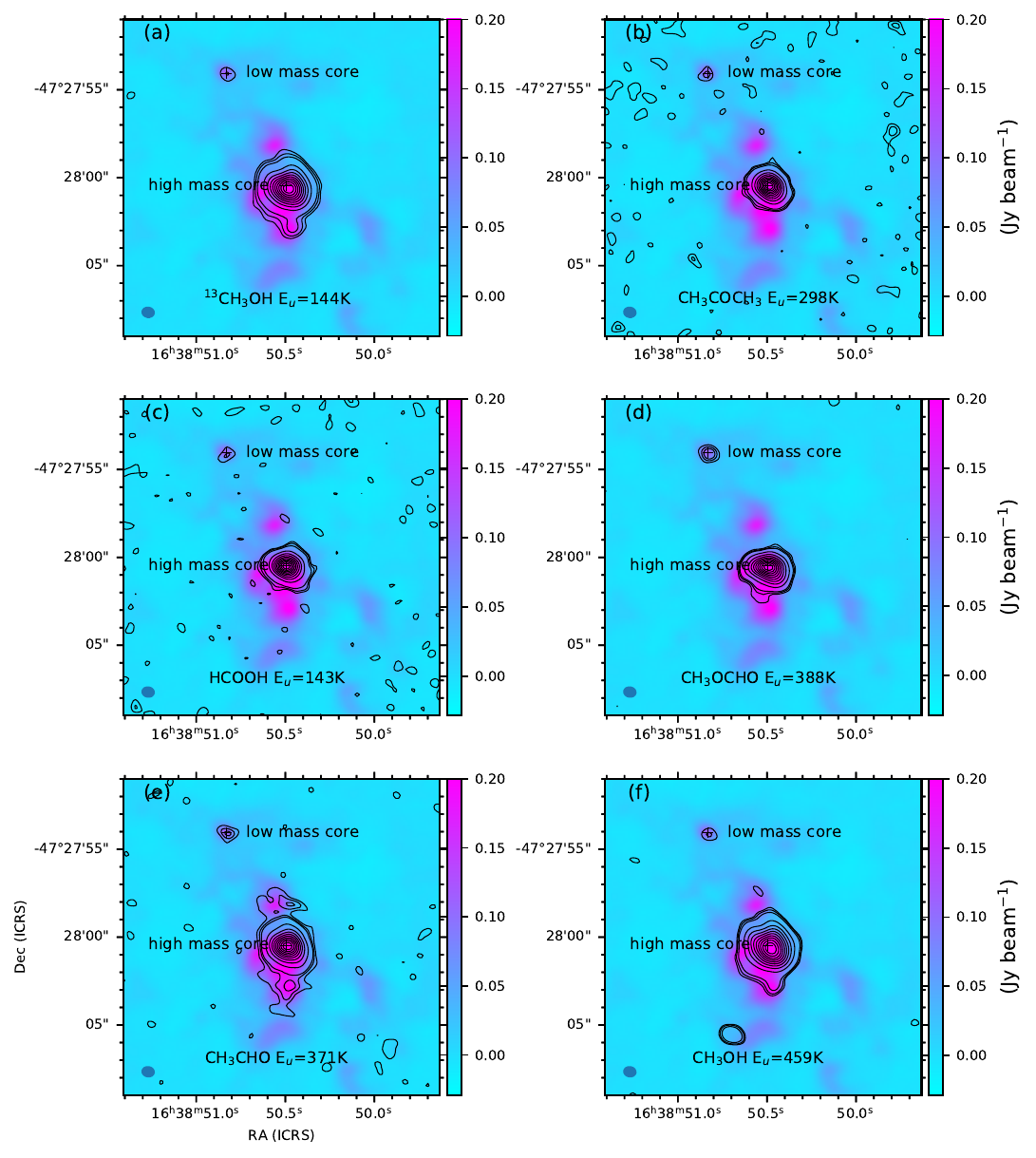}
\caption{Spatial distributions of O-, N- and S-bearing molecules. The black contour levels are [1\%, 2\%, 3\%, 10\%, 20\%, 30\%, 40\%, 50\%, 60\%, 70\%, 80\%, 90\%] $\times$ peak integrated intensities($\rm F_{peak}$). The color grey scale background shows the continuum emission at 0.87\,mm. The synthetic beam for the continuum is indicated in the bottom left corner by the blue ellipse. The value of upper-level energy $E_u$ for each species is shown in the bottom of each panel.} 
\end{figure*}
\addtocounter{figure}{-1}

\begin{figure*}[!ht]
\setcounter{3}{\value{figure}}
\setcounter{figure}{2}
\centering
\includegraphics[scale=0.8]{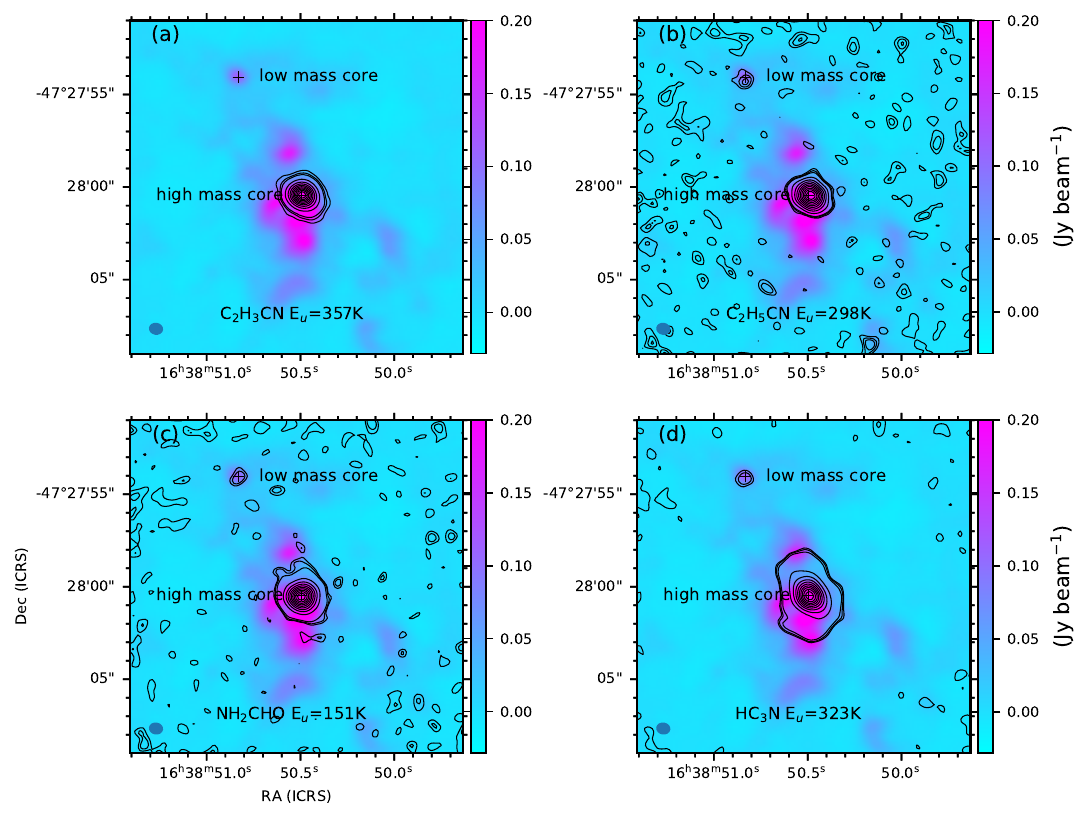}

\includegraphics[scale=0.8]{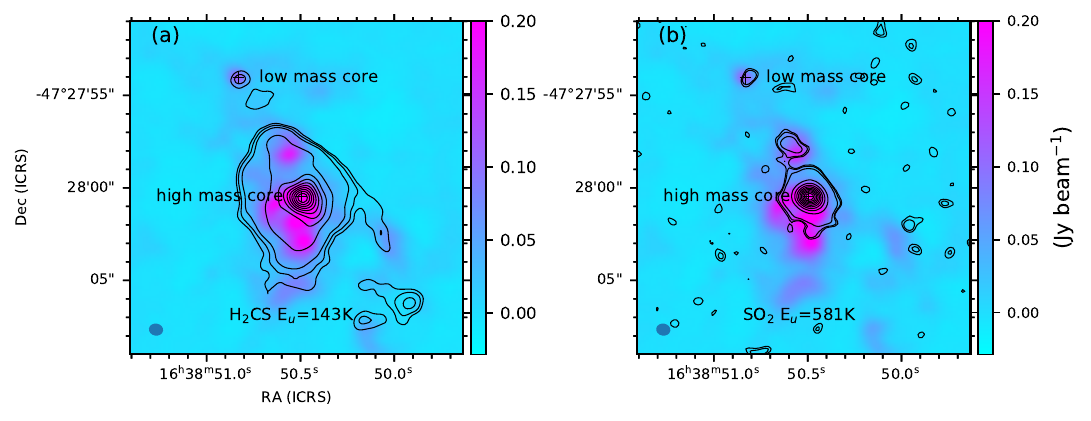}
\caption{(continued)} \label{fig:mom}
\end{figure*}

\section{Discussion}\label{sec:dis}

We modelled the observed spectra using the XCLASS suite. The software contains an interface for the model optimizer package MAGIX (Modelling and Analysis Generic Interface for eXternal numerical codes; \citep{2013A&A...549A..21M}), and finds the best solutions of parameters and provides the corresponding error estimation by using different optimization algorithms (Genetic Algorithm (GA), Levenberg-Marquardt (LM) and Markov Chain Monte Carlo (MCMC) algorithms). The parameters derived from the best-fitting models, the rotational temperature T$_{\rm rot}$ and total column density N$_{\rm tot}$ of the identified molecules are summarized in Table\,\ref{tab:modelfitting} for the high-mass and low-mass cores. The rotational temperatures of molecules range from 93 to 198 K in the high-mass core and from 50 to 130 K in the low-mass core. The column densities range from 2.5$\times$10$^{15}$ cm$^{-2}$ to 4.0$\times$10$^{17}$ cm$^{-2}$ in the high-mass core and 7.0$\times$10$^{13}$ cm$^{-2}$ to 2.0$\times$10$^{16}$ cm$^{-2}$ in the low-mass core. The line widths range from 4.5 to 9.0 km s$^{-1}$ in the high-mass core, and from 1.5 to 3.0 km s$^{-1}$ in the low-mass core. Considering the effect on linewidth due to spectral resolution (0.98 km s$^{-1}$), the deconvolved line width is calculated by FWHM$\rm _{decon}$ = $\rm \sqrt{FWHM^{2}-{\Delta}V^{2}}$, where FWHM is the fitted linewidth convolved with spectral resolution $\rm {\Delta}V^{2}$. Their ranges become 1.13 to 2.83 km s$^{-1}$ in the low-mass core and 4.39 to 8.95 km s$^{-1}$ in the high-mass core.

The fractional abundance of a specific molecule can be obtained by 
$\rm f_{H_2}$ = $\rm N_{tot}$/$\rm N_{H_2}$
($\rm N_{H_2}$=2.3(0.04)$\times$10$^{23}$\,cm$^{-2}$ \\ for the low-mass line-rich core and $\rm N_{H_2}$=1.9(0.2)$\times$10$^{23}$\\\,cm$^{-2}$ for the hot core). The fractional abundances of different species in the two cores are presented in Table\,\ref{tab:modelfitting} according to their abundances in declining order. Together with small mass, abundant line emissions from COMs, high gas temperature, narrow line width, we suggest that the low-mass core in I16351 is a low-mass line-rich core. This is the first detection of a low-mass line-rich core in high-mass star-forming region. The position of the high-mass core is consistent with the hot core position observed by \cite{2022MNRAS.511.3463Q}. CH$_3$OH abundance in the high-mass core is approximately 10$^{-6}$ (Table\,\ref{tab:modelfitting}), and abundances of other COMs ($\sim$ 10$^{-8}$–10$^{-7}$) are reproduced in a hot core model with temperature above 100 K \citep{2013ApJ...765...60G}, which suggest an active hot core chemistry scenario in the high-mass core \citep{2019ApJ...881...57T,2020ApJ...898...54T}.

\begin{figure*}[ht!]
    \centering
    \includegraphics[scale=0.75]{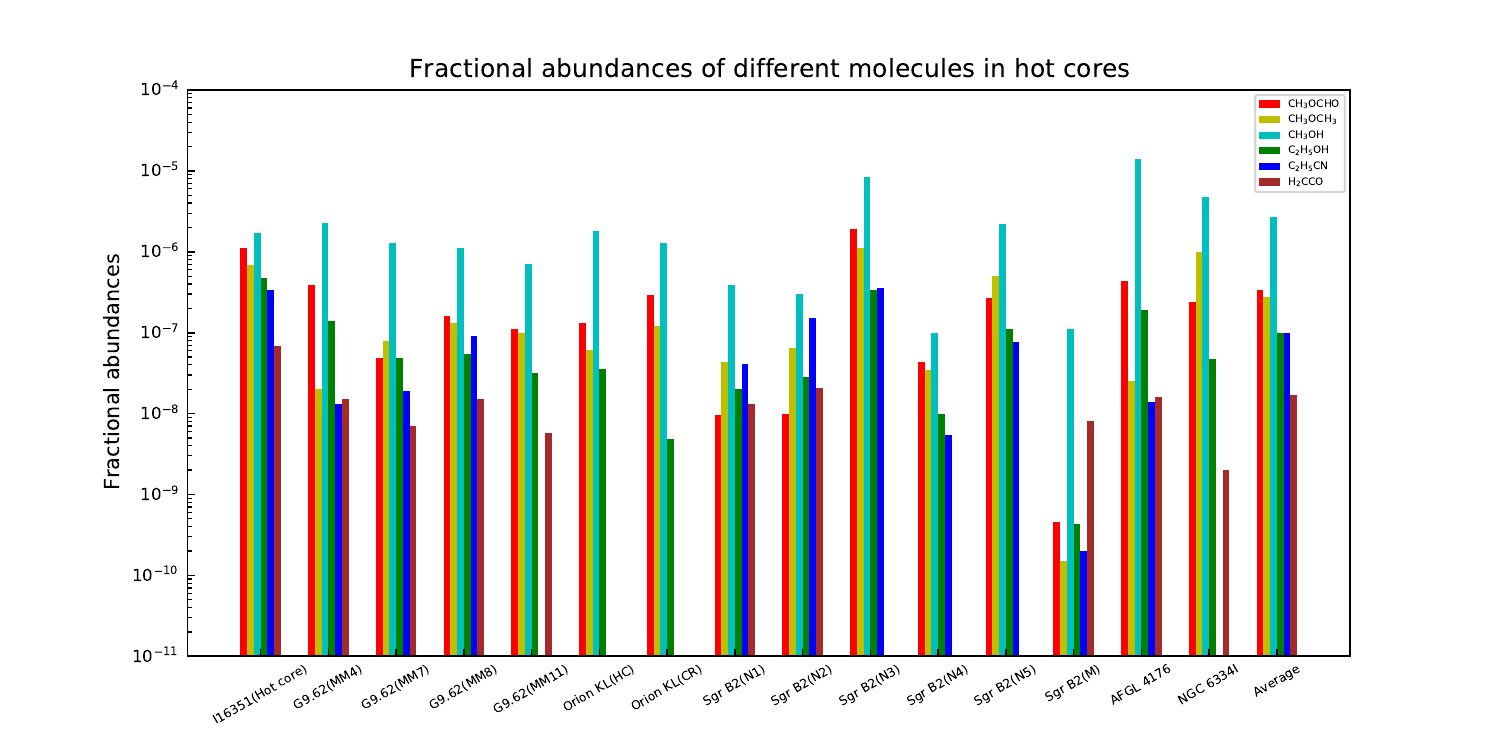}
    \includegraphics[scale=0.75]{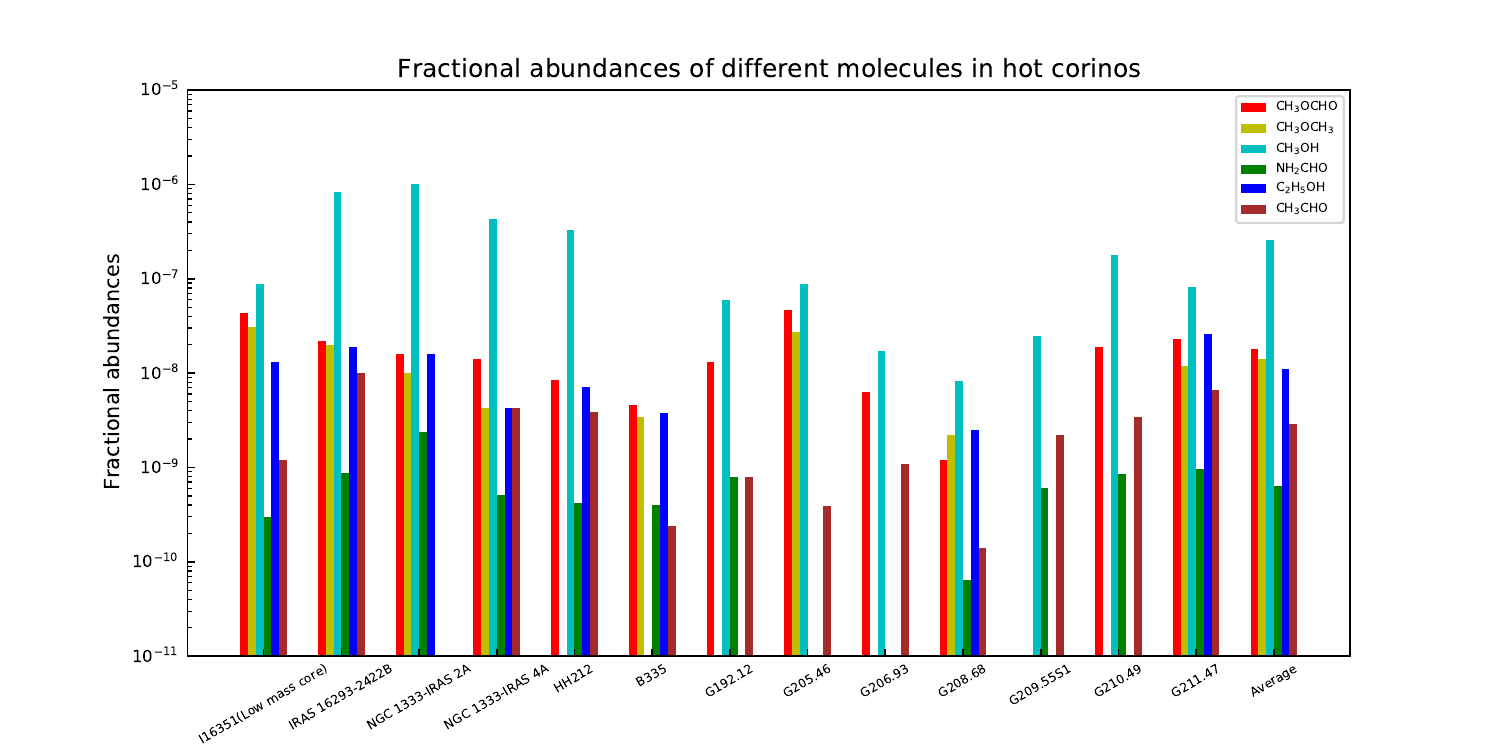}
    \caption{Fractional abundances of different molecules relative to H$_2$ in I16351 and the other hot cores and hot corinos listed in Table\,\ref{tab:molecular abundances} and Table\,\ref{tab:molecular abundances small}.}
    \label{fig:abundance}
\end{figure*}

We have also compiled molecular abundances of CH$_{3}$OCHO, CH$_{3}$OCH$_{3}$, CH$_{3}$OH, C$_{2}$H$_{5}$OH, H$_2$CCO and C$_2$H$_5$CN and CH$_{3}$OCHO, CH$_{3}$OCH$_{3}$, CH$_{3}$OH, C$_{2}$H$_{5}$OH, NH$_{2}$CHO and CH$_3$CHO in some hot cores and hot corinos in Table\,\ref{tab:molecular abundances} and Table\,\ref{tab:molecular abundances small} to compare with our results. Figure\,\ref{fig:abundance} shows the molecular abundance of two cores in I16351 and other hot cores and hot corinos. We find that the fractional abundances of the hot core in I16351 are similar to the average values of other hot cores and the molecular abundances of the low-mass line-rich core in I16351 are similar to the average values of other hot corinos, confirming that a low-mass line-rich core and a hot core are detected in the high-mass star-forming region I16351. CH$_3$OH has high fractional abundances in hot cores and hot corinos. H$_2$CCO and NH$_2$CHO have the lowest fractional abundances in hot cores and hot corinos. Meanwhile, it can be seen from Figure\,\ref{fig:abundance} that abundances of CH$_3$OCH$_3$ are generally higher than those of C$_2$H$_5$OH in hot cores. The abundances of the two species are nearly the same for hot corinos. From Figure\,\ref{fig:abundance} and Tables\,\ref{tab:molecular abundances} and\,\ref{tab:molecular abundances small}, the abundances in the hot core are generally higher than those in the low-mass core by approximately 1$–$2 orders of magnitude. 

\begin{figure*}[ht!]
    \centering
    \includegraphics[scale=0.75]{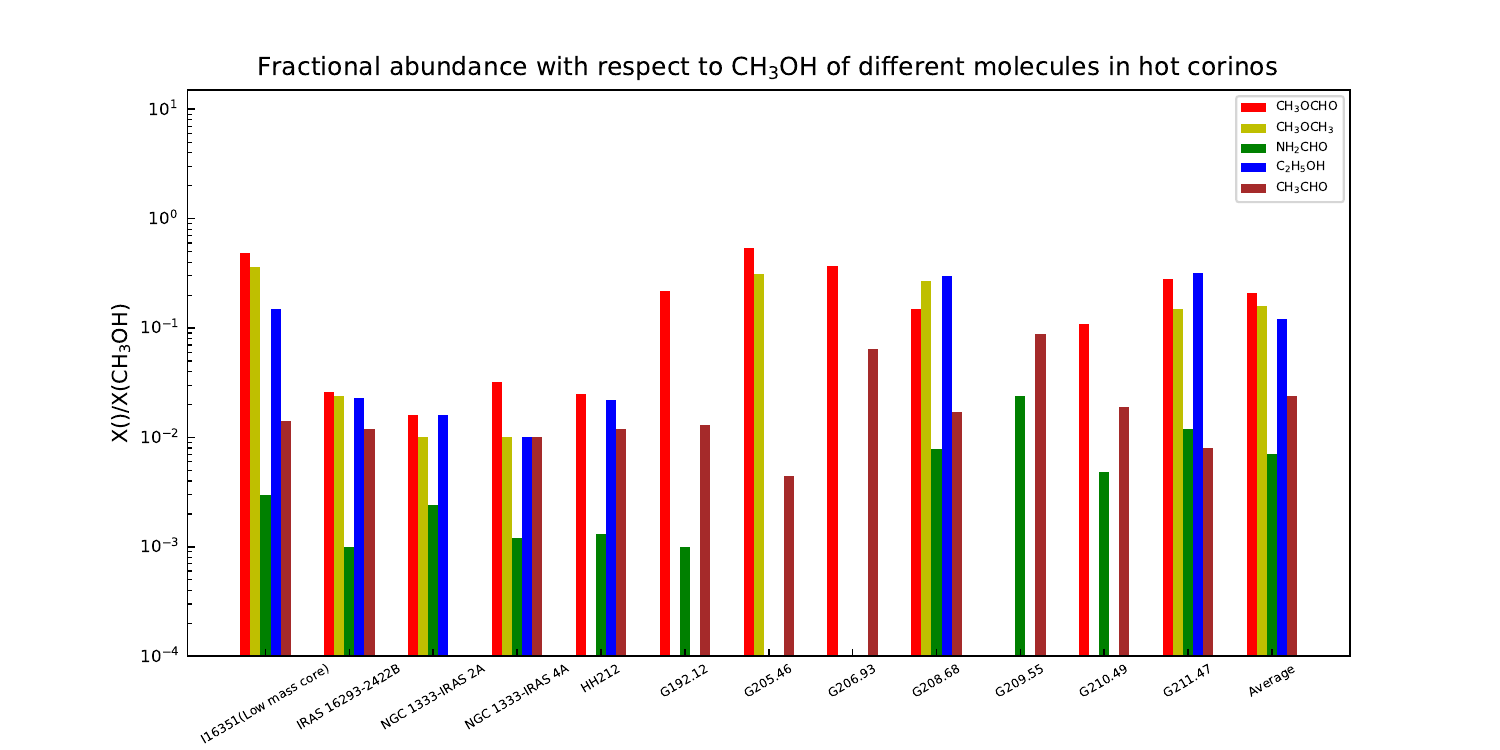}
    \caption{Fractional abundances of different molecules relative to CH$_3$OH in I16351 low-mass line-rich core and the other hot corinos listed in Table\,\ref{tab:molecular abundances small}}.
    \label{fig:abundance ch3oh}
\end{figure*}

Figure\,\ref{fig:abundance ch3oh} compares the molecular abundance with respect to CH$_3$OH of the low-mass core in I16351 with other hot corinos in Table\,\ref{tab:molecular abundances small}. Methyl formate (CH$_3$OCHO) is the most abundant COM with relative abundances of 1.6\%–54\%, followed by dimethyl ether (CH$_3$OCH$_3$) (1\%-36\%) and ethanol (C$_2$H$_5$OH) (1\%–32\%). Other COMs are detected with relative abundances lower than 10\%: Formamide (NH$_2$CHO) has abundances of 0.1\%-2.4\%, while acetaldehyde (CH$_3$CHO) is observed with abundances of 0.44\%–8.8\%. The low-mass core in I16351 has the highest CH$_3$OCH$_3$ abundance among these sources. The relative abundances of COMs in the low-mass core in I16351 are consistent with those in the other hot corinos reported in the literature.

\begin{table*}[ht!]
\centering \caption{Model fitting results of the detected molecules} \label{tab:modelfitting}
\begin{tabular}{cccccc} 
\hline
\hline
Name &     molecule name   &     T$\rm _{rot}$   &  $\rm N_{tot}$   &   $\rm f_{H_2} $$^b$     &  $\rm f_{CH_3OH}$ $ ^c$ \\
     &         &    (K)  &    (cm$^{-2}$)     &     &         \\
\hline
Low-mass line-rich core  & CH$_3$OH; E              & 92(6)   & 3.8 $\times$ 10$^{16}$(0.2) &  &  \\
           & CH$_3$OH; A            & 92(6)   & 2.0 $\times$ 10$^{15}$(0.2) &  &  \\
           & CH$_3$OH$^a$            & 92      & 2.0 $\times$ 10$^{16}$      & 8.7 $\times$ 10$^{-8}$(0.9) & 1 \\
           & CH$_{3}$OCHO v$_{18}$=1 & 90(28)  & 1.1 $\times$ 10$^{16}$(0.4) & 4.8 $\times$ 10$^{-8}$(2.0) & 0.55(0.2)\\
           & CH$_3$OCHO              & 90(9)   & 1.0 $\times$ 
           10$^{16}$(0.1) & 4.3 $\times$ 10$^{-8}$(0.4) & 0.49(0.07)\\
           & $^{13}$CH$_3$OH         & 121(39) & 8.0 $\times$ 10$^{15}$(0.4) & 3.5 $\times$ 10$^{-8}$(0.2) & 0.4(0.05)\\
           & CH$_3$OCH$_3$           & 50(4)   & 7.2 $\times$ 10$^{15}$(0.1) & 3.1 $\times$ 10$^{-8}$(0.06) & 0.36(0.04)\\
           & C$_{2}$H$_{5}$OH        & 130(6)  & 3.0 $\times$ 10$^{15}$(0.6) & 1.3 $\times$ 10$^{-8}$(0.3) & 0.15(0.04)\\
           & aGg'-(CH$_{2}$OH)$_{2}$ & 66(6)   & 2.5 $\times$ 10$^{15}$(0.8) & 1.1 $\times$ 10$^{-8}$(0.3) & 0.13(0.04)\\
           & CH$_3$COCH$_3$          & 93(29)  & 1.3 $\times$ 10$^{15}$(0.2) & 5.7 $\times$ 10$^{-9}$(1.1) & 0.066(0.01)\\
           & H$_2$CS                & 70(5)   & 8.4 $\times$ 10$^{14}$(0.6) & 3.7 $\times$ 10$^{-9}$(0.3) & 0.043(0.006)\\
           & C$_{2}$H$_{5}$CN        & 96(14)  & 5.6 $\times$ 10$^{14}$(1.1) & 2.0 $\times$ 10$^{-9}$(0.4) & 0.023(0.005)\\
           & H$_2$CCO                & 123(19) & 4.8 $\times$ 10$^{14}$(0.8) & 2.1 $\times$ 10$^{-9}$(0.4) & 0.024(0.005)\\
           & CH$_{3}$CHO             & 93(9)   & 2.7 $\times$ 10$^{14}$(0.6) & 1.2 $\times$ 10$^{-9}$(0.2) & 0.014(0.003)\\
           & t-HCOOH                 & 97(4)   & 1.7 $\times$ 10$^{14}$(0.1) & 7.4 $\times$ 10$^{-10}$(0.4)& 0.0085(0.001) \\
           & NH$_{2}$CHO             & 55(18)  & 7.0 $\times$ 10$^{13}$(0.5) & 3.0 $\times$ 10$^{-10}$(0.2)& 0.003(0.0003)\\
\hline
Hot core   & SO$_2$ v$_2$=1          & 198(22) & 4.0 $\times$ 10$^{17}$(2.9) & 2.1 $\times$ 10$^{-6}$(1.5) & 1.24(0.9) \\
           & CH$_3$OH; E              & 142(2)  & 5.0 $\times$ 10$^{17}$(0.1) & & \\
           & CH$_3$OH; A             & 142(2)  & 1.5 $\times$ 10$^{17}$(0.1) &  &  \\
           & CH$_3$OH$^a$            & 142     & 3.3 $\times$ 10$^{17}$      & 1.7 $\times$ 10$^{-6}$(0.2) & 1 \\
           & $^{13}$CH$_3$OH         & 130(37) & 2.2 $\times$ 10$^{17}$(0.1) & 1.2 $\times$ 10$^{-6}$(0.1) & 0.71(0.1) \\
           & CH$_3$OCHO v$_{18}$=1   & 170(31) & 2.0 $\times$ 10$^{17}$(0.9) & 1.1 $\times$ 10$^{-6}$(0.5) & 0.65(0.3) \\
           & CH$_3$OCHO              & 170(10) & 2.0 $\times$ 10$^{17}$(0.1) & 1.1 $\times$ 10$^{-6}$(0.1) & 0.65(0.1)\\
           & CH$_3$OCH$_3$           & 93(5)   & 1.3 $\times$ 10$^{17}$(0.2) & 6.8 $\times$ 10$^{-7}$(1.3) & 0.4(0.09) \\
           & CH$_3$COCH$_3$          & 126(37) & 1.0 $\times$ 10$^{17}$(0.9) & 5.3 $\times$ 10$^{-7}$(4.8) & 0.31(0.28) \\
           & C$_2$H$_5$OH            & 152(19) & 9.0 $\times$ 10$^{16}$(0.5) & 4.7 $\times$ 10$^{-7}$(0.6) & 0.28(0.05)\\
           & C$_2$H$_5$CN            & 149(39) & 6.5 $\times$ 10$^{16}$(0.5) & 3.4 $\times$ 10$^{-7}$(0.4) & 0.2(0.03) \\
           & t-HCOOH                 & 135(49) & 3.7 $\times$ 10$^{16}$(1.5) & 2.0 $\times$ 10$^{-7}$(0.8) & 0.12(0.05) \\
           & C$_2$H$_3$CN v$_{11}$=1 & 171(53) & 3.5 $\times$ 10$^{16}$(1.0) & 1.8 $\times$ 10$^{-7}$(0.5) & 0.11(0.03) \\
           & H$_2$CS                 & 101(1)  & 3.0 $\times$ 10$^{16}$(0.1) & 1.6 $\times$ 10$^{-7}$(0.2) & 0.09(0.02) \\
           & aGg'-(CH$_2$OH)$_2$     & 145(6)  & 2.4 $\times$ 10$^{16}$(1.1) & 1.3 $\times$ 10$^{-7}$(0.6) & 0.076(0.04) \\
           & CH$_3$CHO               & 160(19) & 1.5 $\times$ 10$^{16}$(0.2) & 7.9 $\times$ 10$^{-8}$(1.3) & 0.046(0.009)\\
           & C$_2$H$_3$CN            & 171(8)  & 1.4 $\times$ 10$^{16}$(0.9) & 7.4 $\times$ 10$^{-8}$(4.8) & 0.04(0.03) \\
           & H$_2$CCO               & 159(33) & 1.3 $\times$ 10$^{16}$(0.5) & 6.8 $\times$ 10$^{-8}$(2.7) & 0.04(0.02) \\
           & $^{34}$SO$_2$           & 134(14) & 8.5 $\times$ 10$^{15}$(3.1) & 4.5 $\times$ 10$^{-8}$(1.7) & 0.026(0.01) \\
           & NH$_2$CHO               & 150(39) & 2.5 $\times$ 10$^{15}$(1.3) & 1.3 $\times$ 10$^{-8}$(0.7) & 0.008(0.004) \\
\hline
\end{tabular}
\tablecomments{
$^a$ N$\rm_{CH_3OH}$=[N$\rm_{CH_3OH;~E}$ + N$\rm_{CH_3OH;~A}$]/2, where the capital letter E and A represent different transitions. \\
$^b$ $\rm f_{H_2}$ = $\rm N_{tot}$/ $\rm N_{H_2}$, $\rm N_{H_2}$=2.3(0.04)$\times$10$^{23}$\,cm$^{-2}$ for the low-mass line-rich core and $\rm N_{H_2}$=1.9(0.2)$\times$10$^{23}$\,cm$^{-2}$ for the hot core. \\
$^c$ $\rm f_{CH_3OH}$=N$\rm_{tot}$/N$\rm_{CH_3OH}$.}
\end{table*}

\section{Conclusions}\label{sec:con}

We conducted a comprehensive line survey of complex organic molecules (COMs) in the massive star-forming region I16351 using the high spatial resolution ALMA observations at 870$\,\rm \mu$m wavelength. We have identified distinct chemical differentiation between a hot core and a low-mass core with abundant line emission. The main results are summarized as follows:

\begin{enumerate}
    \item We have identified 12 dust cores from the continuum emission. Of the detected cores, a high-mass core (11.6 M$_{\odot}$) and a low-mass core (1.7 M$_{\odot}$) are the regions rich in line emissions from COMs.
    \item The rotation temperatures and column densities are calculated by the XCLASS software. The rotational temperatures range from 93 to 198\,K for the high-mass core and from 50 to 130\,K for the low-mass core. Line widths in the high-mass core are larger than those of the low-mass one. The high-mass core is considered to be a hot core and the other one is identified as a low-mass line-rich core.
    \item The fractional abundances relative to H$_2$ are derived. We have compared the fractional abundances of COMs in the two cores with other sources, and further confirmed the detection of both a hot core and a low-mass line-rich core found in I16351. It is the first time that a low-mass line-rich core is detected in the massive star-forming region.
    \item The two line-rich cores in I16351 reside in the same region and have different chemical properties, implying that the different chemistry between the two cores are likely influenced by their physical conditions.
\end{enumerate}

\clearpage

\section*{Acknowledgment}

This paper makes use of the following ALMA data: ADS/JAO.ALMA\#2017.1.00545.S. ALMA is a partnership of ESO
(representing its member states), NSF (USA), and NINS (Japan), together with NRC (Canada), MOST and ASIAA (Taiwan), and KASI (Republic of Korea), in cooperation with the Republic of Chile. The Joint ALMA Observatory is operated by ESO, AUI/NRAO, and NAOJ. This work has been supported by National Key R\&D Program of China (No.2022YFA1603101), and by NSFC through the grants No.12033005, No.12073061, No.12122307, and No.12103045. S.-L. Qin thanks the Xinjiang Uygur Autonomous Region of China for their support through the Tianchi Program. MYT acknowledges the support by NSFC through grants No.12203011, and Yunnan provincial Department of Science and Technology through grant No.202101BA070001-261. T. Zhang thanks the student's exchange program of the Collaborative Research Centre 956, funded by the Deutsche Forschungsgemeinschaft (DFG). 

\software{astropy \citep{Astropy2013,Astropy2018}, CASA \citep{McMullin2007CASA}, XCLASS \citep{2017A&A...598A...7M}, MAGIX \citep{2013A&A...549A..21M}}

\appendix

\section{Full line lists}
According to the number of identified emission lines, we arrange the order of molecules in the full list of COM transitions. The list of COM transitions used to derive rotational temperatures and column densities is shown in Table\,\ref{tab:linelists}. Table\,\ref{tab:linelists} in Appendix A shows that the transition, the rest frequencies (GHz), the upper-level energy ($\rm E_u$), and the line strength ($\rm S_\mu^{2}$) of each molecular.

\startlongtable
\begin{deluxetable*}{ccccccc}
\tabletypesize{\footnotesize}
\tablewidth{0pt} 
\tablenum{A1}
\setcounter{table}{0}
\renewcommand{\thetable}{A\arabic{table}}
\centering 
\tablecaption{Properties of Unblended Transitions}
\label{tab:linelists}
\tablehead{
Molecular Name  & Transition   & Frequency & $\rm E_u$ & $\rm S_\mu^2$ & Core name & Note  \\
            &              & (GHz)     & (K)   &   ($\rm Debye^2$)                 
}
\startdata
CH$_3$OCHO v=0  & 11(8,4)-10(7,4)E     & 342.51034  & 81.40352  & 3.46422  & Hot core  &\\
            & 16(13,4)-16(12,5)E   & 342.683424 & 335.27508 & 1.72784  &  Hot core & \\
            & 27(13,14)-27(12,15)E & 342.69288  & 335.27641 & 5.56387  & Hot core/Low-mass line-rich core & \\
            & 27(13,15)-27(12,16)A & 343.72097  & 258.47228 & 5.5622   & Hot core &  \\
            & 27(7,20)-26(7,19)E   & 343.75801  & 258.47556 & 67.17309 & Hot core & \\
            & 28(24,4)-27(24,3)A   & 343.79865  & 589.86175 & 19.94798 & Hot core & \\
                & 28(23,5)-27(23,4)A   & 343.81403  & 589.85602 & 24.43765 & Hot core & \\
                & 28(23,5)-27(23,4)E   & 343.826593 & 589.85    & 24.44059 & Hot core & \\
                & 28(23,6)-27(23,5)E   & 343.835114 & 560.10214 & 24.43829 & Hot core & \\
                & 28(22,6)-27(22,5)A   & 343.854153 & 560.09558 & 28.73324 & Hot core/Low-mass line-rich core & \\
                & 28(22,6)-27(22,5)E   & 343.862096 & 560.09135 & 28.736   & Hot core & \\
                & 28(22,7)-27(22,6)E   & 343.887467 & 531.65761 & 28.73425 & Hot core/Low-mass line-rich core & \\
                & 28(21,8)-27(21,7)A   & 343.887483 & 531.65761 & 32.83771 & Hot core & \\
            & 28(21,7)-27(21,6)A   & 343.921695 & 289.13552 & 32.8377  & Hot core & \\
            & 24(13,11)-24(12,12)A & 345.073057 & 104.42367    & 4.56561  & Hot core & \\
            & 16(6,11)-15(5,10)A &343.958362 & 504.53073     &2.71117     & Hot core & \\
            & 28(20,8)-27(20,7)A & 343.981148 & 504.52218     &36.74693  & Hot core & \\
            & 28(20,9)-27(20,8)E & 343.98245 & 504.52325     &36.74095   & Hot core & \\
            & 28(20,8)-27(20,7)E & 344.322992 & 431.08887     &36.74080  & Hot core & \\
            & 28(17,11)-27(17,10)A & 344.349515 & 431.30336     &47.30097  & Hot core/Low-mass line-rich core & \\
                & 28(17,12)-27(17,11)E &344.515454 & 409.2671  & 47.30336   & Hot core/Low-mass line-rich core &  \\
            & 28(16,12)-27(16,11)E &344.523525 & 409.26331 & 50.42985  & Hot core/Low-mass line-rich core & \\
                & 28(16,12)-27(16,11)E  &344.541314 &409.26057 & 50.42682 & Hot core/Low-mass line-rich core & \\
                & 28(16,13)-27(16,12)E & 344.783597 &388.77694 & 50.43269 & Hot core/Low-mass line-rich core & \\
                & 28(15,14)-27(15,13)E  & 344.957101 & 235.97897 & 53.37525 & Hot core/Low-mass line-rich core & \\
                & 20(13,8)-20(12,9)A  & 344.957127 & 235.97897 & 3.20036 & Hot core/Low-mass line-rich core & \\
                & 20(13,7)-20(12,8)A  & 345.132629 & 224.17841 & 3.20036   & Hot core/Low-mass line-rich core & \\
                & 19(13,6)-19(12,7)A  & 345.132655 &224.17841 & 2.84622    & Hot core & \\
                & 19(13,7)-19(12,8)A & 345.28132  & 212.97193      &2.84622  & Hot core & \\
              & 18(13,5)-18(12,6)A & 345.4058639   & 202.35893      &2.48351  & Hot core & \\
            & 17(13,4)-17(12,5)A & 345.6508346 & 80.31116     &2.11147  & Hot core & \\
          & 9(9,1)-8(8,1)E &345.662771 & 80.33274     &3.89922  & Hot core & \\
            & 9(9,0)-8(8,0)E &345.718662 & 80.31960     &3.89833  & Hot core & \\
            & 9(9,0)-8(8,1)A &342.342185 & 269.49634     &3.89598   & Hot core & \\
          & 30(2,28)-29(3,27)E &342.530119 &269.49054     &10.005  & Hot core/Low-mass line-rich core&  \\
            & 30(2,28)-29(3,27)A &342.35142 & 269.49679     &10.00214  & Hot core & \\
            & 30(3,28)-29(3,27)E &342.358225 & 269.49625  & 78.03662   & Hot core/Low-mass line-rich core &   \\
            & 30(2,28)-29(2,27)E &342.358279 &269.49553 & 78.03333  & Hot core/Low-mass line-rich core & \\
                & 31(1,30)-30(2,29)A  & 343.43526 & 257.07971 & 11.89421 & Hot core/Low-mass line-rich core & \\
                & 28(4,24)-27(4,23)E  & 343.443944 & 257.0807 & 71.57138 & Hot core/Low-mass line-rich core & \\
                & 28(4,24)-27(4,23)A  & 343.541355 & 303.90939 & 71.58443   & Hot core &\\
            & 25(13,13)-25(12,14)E  &343.56185 & 303.91699 & 4.90072    & Hot core &\\
                & 25(13,12)-25(12,13)A & 343.561883  & 303.91699 &4.89913  & Hot core & \\
              & 25(13,13)-25(12,14)A & 344.051371   & 478.72469  &4.89913 & Hot core/Low-mass line-rich core & \\
            & 28(19,9)-27(19,8)A & 344.335357 & 431.08400     &40.45991  & Hot core/Low-mass line-rich core & \\
            &28(17,11)-27(17,10)E & 344.759096 & 388.78295  & 47.29676  & Hot core/Low-mass line-rich core & *\\
            &28(15,13)-27(15,12)A & 344.02926 & 276.09723     &53.37173  & Hot core/Low-mass line-rich core & \\
            &32(1,32)-31( 1,31)E & 345.067795 & 369.64085     &88.1756   & Hot core/Low-mass line-rich core & \\
            &28(14,14)-27(14,13)E & 345.069059 & 369.64307     &56.11927  & Hot core/Low-mass line-rich core &  \\
            &28(14,15)-27(14,14)A & 345.091465 & 369.63767     &56.11927  & Hot core/Low-mass line-rich core & \\
                &28(14,15)-27(14,14)E &345.466962 & 351.85611  & 56.12379   & Hot core/Low-mass line-rich core &   \\
            &28(13,16)-27(13,15)A &345.486602 & 351.85130 & 58.67428  & Hot core/Low-mass line-rich core & \\
                & 28(13,16)-27(13,15)E  &345.974664 &335.43319 & 58.68009 & Hot core/Low-mass line-rich core & \\
                & 28(12,16)-27(12,15)E & 345.985381 &335.43442 & 61.03512 & Hot core/Low-mass line-rich core & \\
                & 28(12,17)-27(12,16)A  & 346.001616 &335.43002 & 61.04559 & Hot core/Low-mass line-rich core & \\
\hline
CH$_3$COCH$_3$ v=0 & 17(17,1)-16(16,1)EA  & 342.4106413 & 146.85179 & 552.43037 & Hot core & \\
            & 40(2,39)-40(0,40)EA  & 342.4856638 & 411.39562 & 41.07562   & Hot core & \\
           & 17(17,0)-16(16,0)EE  & 342.485233 & 147.06155 & 2209.9853    & Hot core & \\
           & 18(14,4)-17(13,5)EA  & 342.596906 & 147.13957 & 3.20786 & Hot core & \\
            & 17(17,1)-16(16,1)EE  & 342.5948875 & 146.9497 & 2210.2712 & Hot core & \\
           & 17(17,1)-16(16,0)AA  & 342.7800345 & 147.04764 & 828.91802    & Hot core & \\
            & 31(4,27)-30(5,26)EA  & 343.338875 & 288.65894 & 881.61793 &Hot core & \\
            & 31(5,27)-30(4,26)EE  & 343.386 & 288.63372 & 3176.311   & Hot core & \\
           & 31(5,27)-30(4,26)AA  & 343.43308 & 288.60878 & 1322.10583    & Hot core & \\
           & 37(27,10)-37(24,13)AA  & 344.46661 & 593.90913 & 21.18788 & Hot core & \\
        & 21(9,12)-21(6,15)AA  & 343.4435689 & 174.6791 & 0.94626   &Hot core/Low-mass line-rich core & \\
    & 32(4,29)-31(3,28)AA  & 344.549874 & 293.69203 & 2389.27358    & Hot core/Low-mass line-rich core & \\
           & 33(2,31)-32(3,30)EE  & 345.639619 & 298.03287 & 592.06425 & Hot core/Low-mass line-rich core & *\\
           & 15(7,8)-14(6,9)EE  & 345.6739538 & 92.93574 & 115.73717    & Hot core/Low-mass line-rich core &  \\
\hline
$^{34}$SO$_2$ v=0  & 8(4,4)-8(3,5)  & 345.1686641 & 71.047 & 9.76929 & Hot core & \\
            & 9(4,6)-9(3,7)  & 345.2856199 & 79.31737 & 11.41398   & Hot core & \\
           & 9(4,6)-9(3,7)  & 345.2856217 & 79.31737 & 11.41398    & Hot core & \\
           & 6(4,2)-6(3,3)  & 345.5530927 & 57.26899 & 6.34216 & Hot core & \\
           & 5(4,2)-5(3,3)  & 345.6512934 &51.76028 & 4.49068    & Hot core & \\
            & 4(4,0)-4(3,1)  & 345.6787871 & 47.1706 & 2.44173   & Hot core & \\
           & 17(4,14)-17(3,15)  & 345.929349 & 178.77039 & 24.49458 & Hot core & \\
           & 19(1,19)-18(0,18)  & 344.5810445 & 167.65667 & 42.23967    & Hot core & \\
            & 13(4,10)-13(3,11)  & 344.8079147 & 121.63281 & 17.92316   & Hot core & \\
            & 15(4,12)-15(3,13)  & 344.9875847 & 148.34497 & 21.20438 & Hot core & \\
           & 11(4,8)-11(3,9)  & 344.9981602 & 98.62644 & 14.6664    & Hot core & \\
            & 7(4,4)-7(3,5)  & 345.5196563 & 63.69764 & 8.08591   & Hot core & \\
\hline
aGg${'}$-(CH$_{2}$OH)$_{2}$ v=0 &34(14.21)v=0-33(14.20)v=1  &342.5799012    &388.78369  &1085.335 & Hot core/Low-mass line-rich core & \\
                        &34(4,31)v=1-33(4,30)v=0    &342.80051910   &297.12558  &990.16715 & Hot core/Low-mass line-rich core & \\
                       &34(3,31)v=1-33(3,30)v=0 & 342.850907 & 297.12209 & 1298.895 & Hot core/Low-mass line-rich core & \\
                       & 27(6,22)v=1-26(5,21)v=1 & 343.0017164 & 205.05936& 47.55677 & Hot core/Low-mass line-rich core &\\
                       & 34(11,23)v=0-33(11,22)v=1  & 343.9543754   & 352.45528 & 909.59335 & Hot core/Low-mass line-rich core &\\
                       & 37(0.37)-v=1-36(0,36)v=1   &343.997789 &286.70541  &317.01833  & Hot core/Low-mass line-rich core & \\
                       & 57(18,40)v=0-57(17,40)v=1  &344.612515 &971.39018  &35.08034& Hot core/Low-mass line-rich core &\\
                    &33(6,27)v=0-32(6,26)v=1    &344.8010266    &299.19716  &1243.89335& Hot core/Low-mass line-rich core &\\
                    & 35(5,31) v=0-34(5,30)v=1  & 344.9464396 & 321.67857 & 1085.81788   & Hot core\\
                  & 33(11,22)v=1-32(11,21)v=0  & 345.7384426 & 335.94816 & 878.0009   & Hot core & \\
\hline
C$_{2}$H$_{5}$OH v=0 &17(2,15)-17(1,17),g-  &344.9501561    &196.49691  &0.31550 &Hot core/Low-mass line-rich core & \\
                    &7(7,1)-6(6,1),g+   &345.17394930   &139.90163  &7.89577&Hot core/Low-mass line-rich core & \\
                    &21(1,21)-20(1,20),g+   &345.2292848    &241.5535   &33.33569&Hot core/Low-mass line-rich core & \\
                    & 21(1,21)-20(1,20),g-  &345.2953553    &246.22196  &32.93056   &Hot core/Low-mass line-rich core & \\
                    &21(0,21)-20(0,20),g-   &345.4081651    &246.20853  &32.93167& Hot core/Low-mass line-rich core & \\
                    &20(3,18)-19(3,17),g+   &345.6485708    &242.48944  &32.28459   &Hot core/Low-mass line-rich core & \\
                    &20(3,18)-19(3,17),g-   &345.6566222    &347.2135   &29.93737   &Hot core/Low-mass line-rich core & \\
                &20(11,9)-19(11,8),g-   &346.0172119    &384.56138  &21.63848&Hot core/Low-mass line-rich core & \\
\hline
C$_2$H$_5$CN v=0   & 43(8,35)-43(7,36)  & 342.622337 & 478.06113 & 34.90162   &Hot core &  \\
           & 15(5,11)-14(4,10)  & 342.6518898 & 79.39247 & 9.20326    & Hot core/Low-mass line-rich core &  \\
            & 15(5,10)-14(4,11)  & 342.6775645 & 79.39255 & 9.20398 & Hot core/Low-mass line-rich core &  \\
            & 38(5,33)-37(5,32) & 343.194574   & 347.4118 & 553.52713 & Hot core/Low-mass line-rich core &  \\
             & 10(6,5)-9(5,4)  & 344.27894 & 63.66672 & 8.97431 & Hot core/Low-mass line-rich core &  \\
            & 39(3,37)-38(3,36)  & 345.921198 & 344.45762 & 573.18734   & Hot core/Low-mass line-rich core &  \\
           & 32(8,25)-32(7,26)  & 345.93802 & 298.17246 & 24.57802    & Hot core/Low-mass line-rich core & * \\
\hline
H$_2$CS v=0  & 10(0,10)-9(0,9)  & 342.9464239 & 90.59115 & 27.19603 & Hot core/Low-mass line-rich core & \\
            & 10(5,5)-9(5,4)  & 343.202331 & 417.55066 & 61.03878   & Hot core/Low-mass line-rich core & \\
           & 10(4,7)-9(4,6)  & 343.3085 & 300.5408 & 22.78476    & Hot core/Low-mass line-rich core & \\
           & 10(2,9)-9(2,8)  & 343.3220819 & 143.34223 & 26.04309 & Hot core/Low-mass line-rich core & \\
            & 10(3,8)-9(3,7)  & 343.4099625 & 209.01721 & 74.06191   & Hot core/Low-mass line-rich core & \\
           & 10(2,8)-9(2,7)  & 343.813168 & 143.37729 & 26.10949    & Hot core/Low-mass line-rich core & *\\
\hline
CH$_3$OCHO v$_{18}$=1  & 28(10,19)-27(10,18)A  & 344.420363 & 492.85974 & 65.12347 & Hot core/Low-mass line-rich core & \\
           & 28(6,23)-27(6,22)A  & 345.148 & 451.66286 & 70.83473 & Hot core/Low-mass line-rich core & \\
            & 28(9,20)-27(9,19)A  & 345.510 & 480.50639 & 66.90806   & Hot core/Low-mass line-rich core & \\
           & 28(12,17)-27(12,16)A  & 343.134119 &521.89351 & 60.99147    & Hot core/Low-mass line-rich core & \\
            & 28(11,17)-27(11,16)A  & 343.66346 & 506.66854 & 63.1497 & Hot core/Low-mass line-rich core & \\
\hline
CH$_3$OCH$_3$ v=0 & 19(0,19)-18(1,18)AE  & 342.60806 & 167.141 & 130.68396   & Hot core/Low-mass line-rich core & \\
           & 17(2,16)-16(1,15)EA  & 343.753306 & 143.69799 & 45.93527    & Hot core/Low-mass line-rich core & \\
           & 19(1,19)-18(0,18)EA  & 344.357816 & 167.17839 & 55.06275 & Hot core/Low-mass line-rich core & \\
            & 38(4,35)-38(5,34)EA  & 344.35804 & 697.06888 & 257.36625   & Hot core & \\
           & 11(3,9)-10(2,8)AA  & 344.518594 & 72.78343 & 29.84891 & Hot core/Low-mass line-rich core & \\
\hline
CH$_3$OH v=0 & 13(1,12)-13(0,13)-+,vt=0  & 342.72983 & 227.4725 & 24.37811 & Hot core/Low-mass line-rich core & \\
           & 18(2,16)-17(3,14)-+,vt=0  & 344.109132 & 419.39854 & 5.31374    & Hot core & \\
            & 18(3,15)-17(4,14)E,vt=0  & 345.91926 & 459.43023 & 22.4084 & Hot core/Low-mass line-rich core & * \\
            & 16(1,15)-15(2,14)A,vt=0  & 345.903916 & 332.6488 & 28.51774   & Hot core/Low-mass line-rich core & \\
           & 19(1,19)-18(2,16)A,vt=0  & 344.443433 & 451.22691 & 23.97181   & Hot core/Low-mass line-rich core & \\
\hline
H$_2$CCO v=0    & 17(0,17)-16(0,16) & 343.17257   & 148.30454 & 34.27746 & Hot core/Low-mass line-rich core & \\
             & 17(4,14)-16(4,13)  & 343.2504 & 356.83923 & 32.38127 & Hot core/Low-mass line-rich core & \\
            & 17(2,16)-16(2,15)  & 343.376133 & 200.53312 & 33.8061   & Hot core/Low-mass line-rich core & \\
           & 17(3,15)-16(3,14)  & 343.384676 & 265.71292 & 99.63928   & Hot core/Low-mass line-rich core &  \\
           & 17(2,15)-16(2,14)  & 343.693935 & 200.60564 & 33.80534   & Hot core/Low-mass line-rich core &  \\
\hline
$^{13}$CH$_3$OH v=0  & 8(-3,6)-9(-2,8) & 344.040629   & 144.4915 & 2.17479 & Hot core/Low-mass line-rich core & *\\
             & 3(3,0)-4(2,2)  & 344.671733 & 61.55273 & 0.24805 & Hot core & \\
            & 2(2,0)-3(1,3)++  & 345.083793 & 44.59627 & 0.30348 &Hot core & \\
           & 4(0,4)-3(-1,3)  & 345.132599 & 35.76015 & 1.54925   & Hot core/Low-mass line-rich core &  \\
\hline
NH$_2$CHO v=0   & 16(3,13)-15(3,12)  & 343.083117 & 165.99525 & 201.84690 &Hot core/Low-mass line-rich core &   \\
            & 17(1,17)-16(1,16)  & 343.1969863 & 151.99629 & 221.24033 & Hot core/Low-mass line-rich core &  \\
            & 17(0,17)-16(0,16)  & 345.181258 & 151.58895 & 221.41017 & Hot core/Low-mass line-rich core & * \\
           & 16(1,15)-15(1,14)  & 345.3253906 & 145.15586 & 207.83422   & Hot core/Low-mass line-rich core & \\
\hline
C$_2$H$_3$CN v=0   & 36(5,31)-35(5,30)  & 342.3755639 & 357.69133 & 1541.37365 &Hot core&  *\\
           & 15(6,9)-16(5,12)  & 342.5893637 & 132.43412 & 4.12349    & Hot core &   \\
            & 36(4,32)-35(4,31)  & 343.4465355 & 338.62854 & 1552.2491 & Hot core &  \\
\hline
t-HCOOH v=0   & 16(1,16)-15(1,15)  & 342.521194 & 143.59095 & 32.16799 & Hot core/Low-mass line-rich core & \\
            & 15(1,14)-14(1,13)  & 343.952343 & 136.28091 & 30.10023   & Hot core/Low-mass line-rich core & \\
           & 16(0,16)-15(0,15)  & 345.030561 & 143.05228 & 31.05632    & Hot core & *\\
\hline
CH$_{3}$C$_{3}$N v=0 & 83(2)-82(2)  &342.6855194    &720.90534  &3742.49011& Low-mass line-rich core &\\
          &83(0)-82(0)& 342.6984005&    690.98043   &3745.47430& Low-mass line-rich core &\\
          &83(3)-82(3)  &342.6694207&   758.3055    &7480.735& Low-mass line-rich core &\\
\hline
CH$_{3}$SH v$_{12}$=1   &16(-1,16)A,vt=1    &344.40893  &480.43037& 2.96268 & Low-mass line-rich core &\\
    &2(-2,1)-1(-1,1)E,vt=1  &344.87788  &325.75082& 0.65398 & Low-mass line-rich core &\\
    &15(1,14)-14(0,14)E,vt=1    &345.336513 &427.10456  &2.5329 & Low-mass line-rich core & \\
\hline
c-H$_{2}$C$_{3}$O v=0   &27(1,27)-26(1,26)  &345.389613 &235.83485  &1554.06816     & Low-mass line-rich core & \\
                        &26(1,25)-25(1,24)  &345.6303058    &233.86764  &1483.9321& Low-mass line-rich core & \\
\hline
SO$_2$ v$_2$=1    & 23(3,21)-23(2,22)  & 342.4359377 & 1040.88125 & 29.57632   & Hot core &    \\
           & 24(2,22)-23(3,21)  & 343.9237569 & 1057.70354 & 19.5591    & Hot core &    \\
\hline
C$_2$H$_3$CN v$_{11}$=1 & 36(7,29)-35(7,28),F=37-36  & 342.928828 & 737.05933 & 511.8907 & Hot core& \\
            & 36(4,33)-35(4,32),F=35-34 & 343.639977   & 667.27558 & 525.44888 & Hot core & \\
\hline
CH$_3$CHO v=0 & 30(2,28)-29(3,27)A, vt=2  & 343.830738 & 827.3158 & 22.24928  & Hot core/Low-mass line-rich core & \\
    & 18(2,17)-17(2,16)A,vt=1  & 343.8330821 & 371.84089 & 225.86166  & Hot core/Low-mass line-rich core & * \\
\hline
CH$_3$C$_4$H v=0 & 85(2)-84(2)&354.8511129 & 781.44479 &266.59178 & Hot core & \\
                & 85(4)-84(4)&345.8116518&876.34192& 270.22466 & Hot core & \\
\hline
HC$_5$N v=0  & J=129-128  & 343.2256445 & 1071.10155 & 7255.3324 & Hot core/Low-mass line-rich core & \\
\hline
SiC$_4$ v=0    & 112-111 & 343.237469   & 931.14787 & 4445.08234 & Hot core & \\
\hline
t-HCOOD v=0  & 40(5,35)-40(4,36)  & 343.97053 & 939.40848 & 2.14618 & Hot core & \\
\hline
CH$_3$OH v$_{12}$=1  & 10(2,9)-11(3,9)E,vt=1  & 344.312267 & 491.90986 & 31.24862 & Hot core/Low-mass line-rich core &   \\
\hline
SO$_2$  v=0 & 34(3,31)-34(2,32)  & 342.7616232 & 581.91876 & 50.73556 & Hot core/Low-mass line-rich core & *\\
\hline
CS v=0      & 7-6  & 342.88285 & 65.8273 & 26.75185   & Hot core/Low-mass line-rich core & \\
\hline
c-C$_2$H$_4$O v=0 &9(5,5)-8(4,4) & 345.68832 &90.40751 &66.52984 & Hot core & \\
\hline
HNCCC v=0 & J=37-36, F=36-37 &345.343 & 314.95191 & 0.86737 & Low-mass line-rich core &  \\
\hline
HC$_3$N v=0   & J=38-37  & 345.60901 & 323.49156 & 529.12907   & Hot core/Low-mass line-rich core & * \\
\hline
NS v=0    & J=15/2-13/2,Ω=1/2  & 345.823288 & 70.79666 & 24.02695   & Hot core/Low-mass line-rich core & \\
          & F=17/2-15/2,l=e    &            &          &            &\\
\hline
\enddata
\tablecomments{The rows marked with * indicate that the transitions of the molecules are used for Figure\,\ref{fig:mom}.}.
\end{deluxetable*}

\section{Fractional Abundance lists}
The fractional abundances of hot cores and hot corinos from other sources reported in literature are listed in Tables\,\ref{tab:molecular abundances} and\,\ref{tab:molecular abundances small}.

\setcounter{table}{0}
\renewcommand{\thetable}{B\arabic{table}}
\centering 
\begin{table*}
\caption{Molecular abundances with respect to H$_2$ in I16351 hot core and other sources.}
\begin{tabular}{cccccccc} 
\hline
\hline
Hot core &     CH$_3$OH   &   CH$_3$OCH$_3$     &  CH$_3$OCHO   & C$_2$H$_5$OH & C$_2$H$_5$CN & H$_2$CCO & Reference \\
&        /[H$_2$] &   /[H$_2$]   & /[H$_2$]     &  /[H$_2$]      &  /[H$_2$]   & /[H$_2$] & \\
\hline
Hot core  & 1.7(-6)& 6.8(-7)& 1.1(-6)& 4.7(-7)& 3.4(-7)& 6.8(-8)& this work\\
G9.62+0.19(MM4)  & 2.3(-6)& 2.0(-7)& 3.9(-7)& 1.4(-7)& 1.3(-8)& 1.5(-8)& 1 \\
G9.62+0.19(MM7)  & 1.3(-6)& 7.8(-8)& 4.8(-8)& 4.8(-8)& 1.9(-8)& 7.0(-9)& 1\\
G9.62+0.19(MM8)  & 1.1(-6)& 1.3(-7)& 1.6(-7)& 5.5(-8)& 9.0(-8)& 1.5(-8)& 1\\
G9.62+0.19(MM11) & 7.1(-7)& 1.0(-7)& 1.1(-7)& 3.2(-8)& -& 5.7(-9)& 1\\
Orion KL(HC)     & 1.8(-6)& 6.1(-8)& 1.3(-7)& 3.6(-8)& - & - & 2  \\
Orion KL(CR)     & 1.3(-6)& 1.2(-7)& 2.9(-7)& 4.9(-9)& - & - & 2  \\
Sgr B2(N1)       & 3.9(-7)& 4.3(-8)& 9.6(-9)& 2.0(-8)& 4.1(-8)& 1.3(-8) & 3  \\
Sgr B2(N2)       & 3.0(-7)& 6.5(-8)& 1.0(-8)& 2.8(-8)& 1.5(-7)& 2.1(-8) & 3\\
Sgr B2(N3)       & 8.3(-6)& 1.1(-6)& 1.9(-6)& 3.4(-7)& 3.6(-7)& - & 4  \\
Sgr B2(N4)       & 9.8(-8)& 3.5(-8)& 4.3(-8)& 9.8(-9)& 5.5(-9)& - & 4  \\
Sgr B2(N5)       & 2.2(-6)& 5.0(-7)& 2.7(-7)& 1.1(-7)& 7.6(-8)& - & 4  \\
Sgr B2(M)        & 1.1(-7)& 1.5(-10)& 4.6(-10)& 4.3(-10)& 2.0(-10)& 8.0(-9) & 3\\
AFGL 4176        & 1.4(-5)& 2.5(-8)& 4.3(-7) & 1.9(-7)& 1.4(-8)& 1.6(-8) & 5 \\
NGC 6334I        & 4.7(-6)& 1.0(-6)& 2.4(-7) & 4.7(-8)& - & 2.0(-9)& 6\\
Average          & 2.7(-6)& 2.8(-7)& 3.4(-7)& 1.0(-7) & 1.0(-7)& 1.7(-8) & \\
\hline
\end{tabular}
\tablecomments{A(B)=A$\times$10$\rm ^B$. 1. \citep{2022MNRAS.512.4419P}. 2.\citep{2018A&A...620L...6T}. 3. \citep{2013A&A...559A..47B}. 4. \citep{2019A&A...628A..27B}. 5. \citep{2019A&A...628A...2B}. 6. \citep{2012A&A...546A..87Z}.}
\label{tab:molecular abundances}
\end{table*}

\begin{table*}
\tablewidth{0pt} 
\linespread{1.2} 
\caption{Molecular abundances with respect to H$_2$ and CH$_3$OH in I16351 hot corino and other sources.} 
\centering
\label{tab:molecular abundances small} 
\scalebox{0.8}{
\begin{tabular}{cccccccc}
\hline
\hline
Hot corino &     CH$_3$OH   &   CH$_3$OCH$_3$     &  CH$_3$OCHO   & C$_{2}$H$_{5}$OH &    NH$_2$CHO  &CH$_3$CHO &Reference \\
&\colhead{/[CH$_3$OH]}{/[H$_2$]} & \colhead{/[CH$_3$OH]}{/[H$_2$]}   & \colhead{/[CH$_3$OH]}{/[H$_2$]}     &  \colhead{/[CH$_3$OH]}{/[H$_2$]}     &  \colhead{/[CH$_3$OH]}{/[H$_2$]}     & \colhead{/[CH$_3$OH]}{/[H$_2$]}  & \\
\hline
Low-mass core & \colhead{1}{8.7(-8)} & \colhead{3.6(-1)}{3.1(-8)} & \colhead{4.9(-1)}{4.3(-8)} & \colhead{1.5(-1)}{1.3(-8)} & \colhead{3.0(-3)}{3.0(-10)} & \colhead{1.4(-2)}{1.2(-9)} & this work\\
IRAS 16293-2422B & \colhead{1}{8.3(-7)} & \colhead{2.4(-2)}{2.0(-8)} & \colhead{2.6(-2)}{2.2(-8)} & \colhead{2.3(-2)}{1.9(-8)} & \colhead{1.0(-3)}{8.7(-10)} & \colhead{1.2(-2)}1.0(-8) &   1,2,3,8  \\
NGC 1333-IRAS 2A & \colhead{1}{1.0(-6)} & \colhead{1.0(-2)}{1.0(-8)} & \colhead{1.6(-2)}{1.6(-8)} & \colhead{1.6(-2)}{1.6(-8)} & \colhead{2.4(-3)}{2.4(-9)} & \colhead{-}{-} &  4  \\
NGC 1333-IRAS 4A & \colhead{1}{4.3(-7)} & \colhead{1.0(-2)}{4.3(-9)} & \colhead{3.2(-2)}{1.4(-8)} & \colhead{1.0(-2)}{4.3(-9)} & \colhead{1.2(-3)}{5.1(-10)} & \colhead{1.0(-2)}{4.3(-9)} & 4,9  \\
HH212 & \colhead{1}{3.3(-7)} & \colhead{-}{-} & \colhead{2.5(-2)}{8.4(-9)} & \colhead{2.2(-2)}{7.1(-9)} & \colhead{1.3(-3)}{4.2(-10)} & \colhead{1.2(-2)}{3.9(-9)} & 5\\
B335 & \colhead{-}{-} & \colhead{-}{3.4(-9)} & \colhead{-}{4.6(-9)} & \colhead{-}{3.8(-9)} & \colhead{-}{4.0(-10)} & \colhead{-}{2.4(-10)} & 6\\
G192.12-11.10 & \colhead{1}{6.0(-8)} & \colhead{-}{-} & \colhead{2.2(-1)}{1.3(-8)} & \colhead{-}{-} & \colhead{1.3(-2)}{7.9(-10)} & \colhead{1.3(-2)}{7.9(-10)} & 7\\
G205.46-14.56S1-A & \colhead{1}{8.7(-8)} & \colhead{3.1(-1)}{2.7(-8)} & \colhead{5.4(-1)}{4.7(-8)} & \colhead{-}{-} & \colhead{-}{-} & \colhead{4.4(-3)}{3.9(-10)} & 7\\
G206.93-16.61W2 & \colhead{1}{1.7(-8)} & \colhead{-}{-} & \colhead{3.7(-1)}{6.3(-9)} & \colhead{-}{-} & \colhead{-}{-} & \colhead{6.5(-2)}{1.1(-9)} & 7\\
G208.68–19.20N1 & \colhead{1}{8.2(-9)} & \colhead{2.7(-1)}{2.2(-9)} & \colhead{1.5(-1)}{1.2(-9)} & \colhead{3.0(-1)}{2.5(-9)} & \colhead{7.8(-3)}{6.4(-11)} & \colhead{1.7(-2)}{1.4(-10)} & 7\\
G209.55–19.68S1 & \colhead{1}{2.5(-8)} & \colhead{-}{-} & \colhead{-}{-} & \colhead{-}{-} & \colhead{2.4(-2)}{6.0(-10)} & \colhead{8.8(-2)}{2.2(-9)} & 7\\
G210.49–19.79W–A & \colhead{1}{1.8(-7)} & \colhead{-}{-} & \colhead{1.1(-1)}{1.9(-8)} & \colhead{-}{-} & \colhead{4.8(-3)}{8.6(-10)} & \colhead{1.9(-2)}{3.4(-9)} & 7\\
G211.47–19.27S & \colhead{1}{8.2(-8)} & \colhead{1.5(-1)}{1.2(-8)} & \colhead{2.8(-1)}{2.3(-8)} & \colhead{3.2(-1)}{2.6(-8)} & \colhead{1.2(-2)}{9.7(-10)} & \colhead{8.0(-3)}{6.6(-9)} & 7\\
Average & \colhead{1}{2.6(-7)}&\colhead{1.6(-1)}{1.4(-8)}&\colhead{2.1(-1)}{1.8(-8)}&\colhead{1.2(-1)}{1.1(-8)}&\colhead{7.0(-3)}{6.3(-10)}&\colhead{2.4(-2)}{2.9(-9)} & \\
\hline
\end{tabular}
}
\tablecomments{A(B)=A$\times$10$\rm ^B$. 1.\citep{2018A&A...620A.170J}. 2.\citep{2020ARA&A..58..727J}. 3.\citep{2019MNRAS.490...50D}. 4.\citep{2015ApJ...804...81T}. 5.\citep{2017ApJ...843...27L,2019ApJ...879..101L}. 6.\citep{2016ApJ...830L..37I}. 7.\citep{2022ApJ...927..218H}. 8.\citep{2016A&A...590L...6C}. 9.\citep{2017A&A...606A.121L}}
\end{table*}

\clearpage
\bibliographystyle{aasjournal}
\bibliography{reference}{}

\end{document}